\documentclass[multphys,vecphys]{svmult}

\usepackage{makeidx}         
\usepackage{graphicx}        
\usepackage{multicol}        
\usepackage[bottom]{footmisc}

\begin{document}


\newcommand{\vnhat}{\hat{\vec{n}}}
\newcommand{\vkhat}{\hat{\vec{k}}}
\newcommand{\ve}{\vec{e}}
\newcommand{\bE}{\vec{E}}
\newcommand{\vk}{\vec{k}}
\newcommand{\vl}{\vec{l}}
\newcommand{\vm}{\vec{m}}
\newcommand{\vn}{\vec{n}}
\newcommand{\vv}{\vec{v}}
\newcommand{\vvb}{\vec{v}_{\mathrm{b}}}
\newcommand{\vvgam}{\vec{v}_\gamma}
\newcommand{\vp}{\vec{p}}
\newcommand{\vx}{\vec{x}}
\newcommand{\vtheta}{\vec{\theta}}
\newcommand{\vphi}{\vec{\phi}}
\newcommand{\clp}{\mathcal{P}}
\newcommand{\vd}{\vec{d}}
\newcommand{\vs}{\vec{s}}
\newcommand{\mA}{\tens{A}}
\newcommand{\mN}{\tens{N}}
\newcommand{\mS}{\tens{S}}
\newcommand{\mC}{\tens{C}}
\newcommand{\mB}{\tens{B}}

\newcommand{\nel}{n_{\mathrm{e}}}
\newcommand{\sigT}{\sigma_{\mathrm{T}}}
\newcommand{\velb}{v_{\mathrm{b}}}
\newcommand{\velgam}{v_\gamma}
\newcommand{\dotvelb}{\dot{v}_{\mathrm{b}}}
\renewcommand{\Theta}{\varTheta}
\renewcommand{\Delta}{\varDelta}

\newcommand{\apj}{Astrophysical.~J.}
\newcommand{\apjs}{Astrophysical.~J.~Suppl.}
\newcommand{\apjl}{Astrophysical.~J.~Lett.}
\newcommand{\prd}{Phys.~Rev.~D}
\newcommand{\prl}{Phys.~Rev.~Lett.}
\newcommand{\nat}{Nature}
\newcommand{\mnras}{Mon.~Not.~R.~Astron.~Soc.}
\newcommand{\aap}{Astron.~Astrophys.}
\newcommand{\aaps}{Astron.~Astrophys.~Suppl.}

\title*{Cosmic microwave background polarization analysis}
\author{Anthony Challinor}
\institute{Astrophysics Group, Cavendish Laboratory, Madingley Road,\\
Cambridge, CB3 0HE,  U.K. \\
\texttt{a.d.challinor@mrao.cam.ac.uk}}
%
%


\maketitle

\begin{abstract}
With polarization of the cosmic microwave background (CMB) now detected,
and confirmed by several independent experiments, the next goal is to
characterise accurately its statistical properties.
In these lecture notes we review the physical motivation
for pursuing CMB polarization, and the basic statistical properties of the
polarization fields. We then discuss some of the key aspects of the analysis of
CMB polarization data, focusing on the additional complications that arise
compared to temperature data due to the tensor character of the
polarization field.
\end{abstract}

\section{Introduction}
\label{sec:intro}

The observed statistical properties of the temperature anisotropies
in the cosmic microwave background (CMB) have played a key role in
shaping our understanding of the large-scale properties of the Universe.
The angular power spectrum of the anisotropies, quantifying the contribution
to the variance of the anisotropies as a function
of angular scale, has
now been measured exquisitely over two decades of scale by the
Wilkinson Microwave Anisotropy Probe (WMAP) satellite~\cite{adc:bennett03}. 
A number of ground and balloon-borne experiments extend the power spectrum
measurements by a further decade in scale but with somewhat less precision
than has been attained on the larger scales (see~\cite{adc:bond03} for a recent
review). Current CMB data is remarkably
consistent with predictions based on the passive evolution of
density perturbations, with adiabatic initial conditions
drawn from a Gaussian distribution with an almost-scale-invariant
power spectrum, in a spatially-flat universe. Such primordial fluctuations,
and the observed flat geometry of space, are natural outcomes of the
simplest models of inflation in the early universe,
e.g.~\cite{adc:guth81}.

In addition to having angular variations in the total intensity,
with r.m.s.\ $\sim 120 \,\mu\mathrm{K}$, the
CMB is also partially linearly polarized with r.m.s.\ $\sim 6\,\mu\mathrm{K}$.
Polarization is generated by Thomson scattering of CMB photons around
the time of recombination, once scattering becomes sufficiently rare to
allow the development of anisotropies in the CMB intensity. Polarization
is further generated at late times by re-scattering once the Universe
reionizes. Despite being predicted shortly after the
discovery of the CMB~\cite{adc:rees68},
linear polarization wasn't detected until
2002 by the Degree Angular Scale Interferometer (DASI)
team~\cite{adc:kovac02}. It has now been measured by two further
groups~\cite{adc:readhead04,adc:barkats05},
and also detected via its correlation with the temperature
anisotropies in the first-year WMAP data~\cite{adc:kogut03}.
The direct measurements are currently
much less precise than those for the temperature anisotropies, but
they are fully consistent with predictions in models favoured by the
temperature data. Future, precise measurements of CMB polarization will
be very valuable, allowing accurate determination of several cosmological
parameters that are almost degenerate in their effect on the temperature
anisotropies~\cite{adc:zaldarriaga97inf}.
The biggest winners will be the reionization
history~\cite{adc:zaldarriaga97reion},
with the temperature--polarization correlation observed by
WMAP already constraining the optical depth back to reionization to
be $\sim 0.15$~\cite{adc:kogut03}, and constraints on any gravitational-wave
background~\cite{adc:seljak97,adc:kamionkowski97}
and sub-dominant isocurvature modes~\cite{adc:bucher01}.

As the accuracy and size of CMB datasets have grown, so have the demands
on the techniques used to process the data. A standard set of compression
and cleaning steps have emerged that ultimately distill $O(10^{10})$
time-ordered
observations to several hundred power spectrum measurements, and only around
10 cosmological parameters and their errors (see e.g.~\cite{adc:bond99}
for a review.)
Although the detail of some of these
steps are instrument-specific, most analyses can be regarded as (approximate)
variants on a small number of generic algorithms. Three of the most important
steps are reviewed in detail elsewhere in this volume: for map making and
power spectrum estimation see the contribution by Borrill; for (astrophysical)
component separation and foreground removal see that by Delabrouille.
Most of the steps in
the analysis of CMB polarization data are straightforward generalisations
of those for the temperature, but some new issues do arise mainly due to the
tensorial nature of the polarization field. Furthermore, the small
amplitude of the polarization signal demands that even more careful attention
be paid to potential systematic effects in the analysis of polarization
data than for the temperature anisotropies. Inevitably, this makes polarization
data processing more instrument-specific and was the reason that the
WMAP team did not release polarization maps with their first-year data.

The purpose of these lecture notes is to describe, in outline, the scientific
motivation for pursuing CMB polarization, and discuss some of the key
steps in the analysis of polarization data. We shall focus on those
parts of the analysis that are complicated by the different geometric character
of linear polarization compared to temperature anisotropies. To keep
the discussion general, we will not have much to say on detailed
instrument-specific parts of the analysis, although these are clearly crucial
to the success of any given observation. We start in Sect.~\ref{adc:secstats}
with a discussion of the statistics of the CMB polarization fields and
their representation in terms of orthonormal basis functions on the sphere.
Section~\ref{adc:secphysics} reviews the basic physics of CMB polarization,
and the current detections and best
upper limits on CMB polarization are discussed in
Sect.~\ref{adc:secmeasure}. What we have already learnt from current
polarization measurements, and what we may learn in the future, is
described in Sect.~\ref{adc:secmotivation}.
The analysis of polarization data is then treated in
Sect.~\ref{adc:secanalysis},
where discussions of map-making, astrophysical component separation,
$E$- and $B$-mode separation, power spectrum estimation and
non-Gaussian lensing issues can be found.

\section{Statistics of CMB Polarization}
\label{adc:secstats}

The polarization on the sky along some line of sight $\vnhat$ is 
conveniently described in terms of Stokes specific brightness parameters
$I(\vnhat)$, $Q(\vnhat)$, $U(\vnhat)$ and $V(\vnhat)$. For the CMB
the total intensity $I(\vnhat)$ has a Planck spectrum along any line
of sight, but with angular temperature variations $\Delta T(\vnhat)$
at the $10^{-5}$ level on a 2.725-K background. In linear theory the
anisotropic contribution to $I(\vnhat)$ is thus proportional to
$\Delta T(\vnhat)$ and has a spectrum given by the derivative of
the Planck function (with respect to temperature). The parameters
$Q$ and $U$ describe linear polarization, and inherit the spectrum of the
anisotropic part of $I(\vnhat)$ since CMB polarization is produced by
frequency-independent Thomson scattering. The parameter $V$ describes
circular polarization and is expected to vanish for the CMB.

For a line of sight $\vnhat$, the radiation is propagating along
$-\vnhat$. If we introduce a pair of orthogonal directions in the surface of
the sphere, and call these $x$ and $y$, such that $x$, $y$ and $-\vnhat$
form a right-handed basis, then the linear Stokes parameter $Q$ can be defined
operationally as the difference between the intensities transmitted by perfect
polarizers oriented along $x$ and $y$. Similarly, $U$ is the difference
in intensities if the polarizers are rotated by $45^\circ$. We shall adopt the
convention that in a spherical-polar coordinate system, the local
$x$-axis is along the direction of decreasing polar angle $\theta$ (i.e.\
north) and the $y$-axis is along the direction of increasing
azimuthal angle $\phi$ (i.e.\ east). This is in accordance with the IAU
recommendations, but differs from e.g.~\cite{adc:seljak97} and the latest
version (1.2) of the HEALPix
package\footnote{http://www.eso.org/science/healpix/}.
If the local $x$- and $y$-axes are rotated
through an angle $\psi$ in a right-handed sense about the propagation
direction, $Q$ and $U$ transform as
\begin{eqnarray}
Q & \rightarrow & Q \cos 2\psi + U \sin 2\psi \; ,\nonumber \\
U & \rightarrow & U \cos 2\psi - Q \sin 2\psi \label{adc:eq1} \; .
\end{eqnarray}
These show that $Q$ and $U$ properly form the components (in an orthonormal
basis) of a rank-2, symmetric trace-free tensor
$\clp^{ab}(\vnhat)$ that is transverse to the line of sight:
\begin{equation}
\clp^{ab}(\vnhat) = \frac{1}{2}[Q(\hat{\vtheta}\otimes \hat{\vtheta}
- \hat{\vphi}\otimes \hat{\vphi}) - U(\hat{\vtheta}\otimes \hat{\vphi} +
\hat{\vphi}\otimes \hat{\vtheta})] \; .
\label{adc:eq2}
\end{equation}
This linear polarization tensor is proportional to the correlation tensor
(at zero lag) of the electric field components for the radiation propagating
along $-\vnhat$.

\begin{figure}[t!]
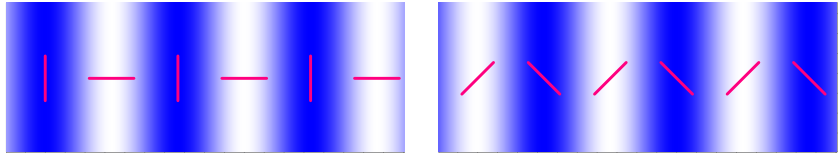

\begin{center}
\includegraphics[angle=-90,width=0.45\textwidth]{adc_f1b.ps}
\quad
\includegraphics[angle=-90,width=0.45\textwidth]{adc_f1a.ps}
\end{center}
\caption{Polarization patterns for a pure-electric mode (left) and
pure-magnetic mode (right) on a small patch of the sky for potentials that
are locally Fourier modes. The shading denotes the amplitude of the potential.
For the electric pattern the polarization is aligned with or perpendicular to
the Fourier wavevector depending on the sign of the potential; for the
magnetic pattern the polarization is at $45^\circ$. 
}
\label{adc:fig1} 
\end{figure}

Any two-dimensional tensor with the symmetries of $\clp_{ab}$ can be derived
from two scalar fields, denoted $P_E$ and $P_B$,
via~\cite{adc:seljak97,adc:kamionkowski97}:
\begin{equation}
\mathcal{P}_{ab} = \nabla_{\langle a} \nabla_{b\rangle} P_E +
\epsilon^c{}_{\langle a}\nabla_{b\rangle} \nabla_c P_B \; ,
\label{adc:eq3}
\end{equation}
where angle brackets denote the symmetric, trace-free part of the enclosed
indices, $\nabla_a$ is the covariant derivative on the sphere, and
$\epsilon_{ab}$ is the alternating tensor. The part derived from
$P_E$ is the \emph{electric} (or gradient) part and that from $P_B$ the
\emph{magnetic} (or curl) part of $\clp_{ab}$. The divergence
$\nabla^a \mathcal{P}_{ab}$ is a pure gradient if
$P_B = 0$, and a curl if $P_E = 0$. The
decomposition~(\ref{adc:eq3}) is analogous to writing a vector field
as a gradient and a divergence free vector, since in two dimensions the
latter is always of the form ${\epsilon_a}^b \nabla_b \chi$. To gain some
intuition for $E$ modes and $B$ modes, consider the case where
$P_E$ and $P_B$ behave locally like a plane wave across a small patch of the
sky (that can accurately be treated as flat). The electric and magnetic
contributions to the polarization are depicted in Fig.~\ref{adc:fig1}.
Quite generally, for a given potential $P$, the magnetic polarization
$\clp_{ab}^B \equiv \epsilon^c{}_{\langle a}\nabla_{b\rangle} \nabla_c P$
generated from $P$ is related to the electric polarization $\clp^E_{ab} \equiv
\nabla_{\langle a} \nabla_{b\rangle} P$ by $\clp^B_{ab} = {\epsilon^c}_{(a}
\clp^E_{b)c}$ which is a right-handed rotation of $\clp^E_{ab}$
by $45^\circ$ about the line of sight.

The fields $P_E$ and $P_B$ are simple scalars that can be expanded in spherical
harmonics with $l \geq 2$ as
\begin{equation}
P_E(\vnhat) = \sum_{lm} \sqrt{\frac{(l-2)!}{(l+2)!}} E_{lm} Y_{lm}(\vnhat)\; ,
\quad P_B(\vnhat) = \sum_{lm} \sqrt{\frac{(l-2)!}{(l+2)!}} B_{lm}
Y_{lm}(\vnhat)\;.
\label{adc:eq3b}
\end{equation}
(The normalisation is conventional.)
The polarization tensor can then
be written as
\begin{equation}
\clp_{ab}(\vnhat) = \frac{1}{\sqrt{2}} \sum_{lm} E_{lm} Y^E_{(lm)ab}
+ B_{lm} Y^B_{(lm)ab} \; ,
\label{adc:eq4}
\end{equation}
where the trace-free, symmetric tensors
\begin{eqnarray}
Y^E_{(lm)ab} & \equiv & \sqrt{\frac{2(l-2)!}{(l+2)!}} \nabla_{\langle a}
\nabla_{b \rangle} Y_{lm} \; ,\nonumber \\
Y^B_{(lm)ab} & \equiv & \sqrt{\frac{2(l-2)!}{(l+2)!}} \epsilon^c{}_{(a}
\nabla_{b)}\nabla_c Y_{lm} \; , \label{adc:eq5}
\end{eqnarray}
form orthonormal tensor bases for electric and magnetic polarization
respectively. The electric and magnetic multipoles, $E_{lm}$ and $B_{lm}$,
can be extracted directly from the polarization tensor by contraction
with the appropriate tensor harmonic followed by integration over the sphere.
For example,
\begin{equation}
E_{lm} = \sqrt{2} \int \clp^{ab} Y^{E*}_{(lm)ab}\, \D\vnhat \; .
\label{adc:eq6} 
\end{equation}
Methods of recovering $E$ and $B$ modes when observations cover only
a limited part of the sky are discussed later in
Sect.~\ref{adc:secanalysis}. Reality of $P_E$ and $P_B$ demands that
$E_{lm}^* = (-1)^m E_{l\, -m}$, and similarly for $B_{lm}$.

We end this subsection by noting an alternative formalism for expressing
symmetric, trace-free tensors on the sphere: the spin-weighted
formalism~\cite{adc:goldberg67}, first introduced to
CMB physics in~\cite{adc:seljak97}.
The complex polarization $P \equiv Q + \I U$
is scaled by a phase factor $e^{-2\I\psi}$ under a right-handed rotation
of the local $x$- and $y$-directions about the propagation direction,
and so is defined to have spin -2. The appropriate basis functions
for expanding functions of definite spin on the sphere are the
spin-weighted spherical harmonics, ${}_{\pm 2}Y_{lm}(\vnhat)$,
which are related to the components of the tensor harmonics,
$Y^E_{(lm)ab}$ and $Y^B_{(lm)ab}$, in a null diad:
\begin{eqnarray}
Y^E_{(lm)ab} &=& \frac{1}{\sqrt{2}}({}_{-2} Y_{lm} m_a m_b
 + {}_{2} Y_{lm} m^*_a m^*_b ) \; , \nonumber \\
Y^B_{(lm)ab} &=& \frac{1}{\I\sqrt{2}}({}_{-2} Y_{lm} m_a m_b
- {}_{2} Y_{lm} m^*_a m^*_b ) \; . \label{adc:eq7}
\end{eqnarray}
Here the null (complex) vector $\vm \equiv (\hat{\vtheta} + \I \hat{\vphi})
/ \sqrt{2}$ satisfies $\vm \cdot \vm = 0$ and $\vm \cdot \vm^* = 1$.
With $Q$ and $U$ expressed on the
north--east basis of a spherical-polar coordinate system, the multipole
expansion of the complex polarization is
\begin{equation}
(Q\pm \I U)(\vnhat) = \sum_{lm} (E_{lm} \mp \I B_{lm}) {}_{\mp 2}
Y_{lm}(\vnhat) \; .
\label{adc:eq8}
\end{equation}
The electric and magnetic multipoles can be extracted using the
orthonormality of the spin-weighted harmonics, e.g.\
\begin{equation}
E_{lm} \mp \I B_{lm} = \int (Q \pm \I U) {}_{\mp 2}Y_{lm}^*
\, \D \vnhat \; .
\label{adc:eq9}
\end{equation}

\subsection{Polarization Power Spectra}

The decomposition of the polarization field into electric and
magnetic parts is invariant under rotations, and the electric and
magnetic multipoles at fixed $l$ transform irreducibly under rotations,
i.e.\ $E_{lm} \rightarrow D^l_{mm'} E_{lm'}$ where $D^l_{mm'}$ are
the Wigner functions (see e.g.~\cite{adc:varshalovich}).
Under the operation of parity,
$(Q\pm \I U)(\vnhat) \rightarrow (Q\mp \I U)(-\vnhat)$ so that
$E_{lm} \rightarrow (-1)^l E_{lm}$ (electric parity) while
$B_{lm} \rightarrow (-1)^{l+1} B_{lm}$ (magnetic parity). These
transformations ensure that the potential $P_E$ is a scalar under parity,
$P_E(\vnhat) \rightarrow P_E(-\vnhat)$, but $P_B$ is a pseudo-scalar,
$P_B(\vnhat) \rightarrow -P_B(-\vnhat)$. The temperature
anisotropies $\Delta T(\vnhat)$ are a scalar function and so can be
expanded in spherical harmonics in the usual way: $\Delta T(\vnhat) =
\sum_{lm} T_{lm} Y_{lm}$. The multipoles $T_{lm}$ have electric parity.

\begin{figure}[t!]
\begin{center}
\includegraphics[height=10cm,angle=-90]{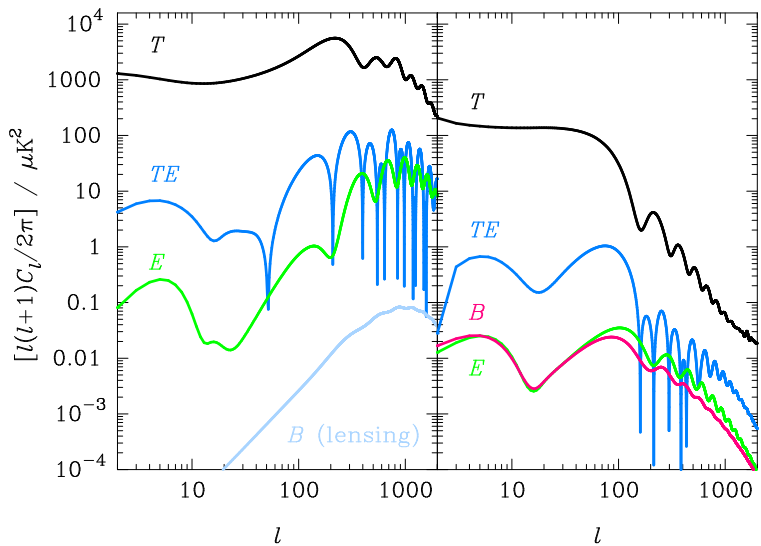}
\end{center}
\caption{Power spectra produced by adiabatic scalar perturbations (left) and
tensor perturbations (right) for a tensor-to-scalar ratio $r=0.38$.
}
\label{adc:fig2} 
\end{figure}

In the absence of parity-violating interactions, the homogeneous and
isotropic background universe on which cosmological perturbations are
assumed to propagate can support an ensemble of fluctuations that is invariant
under parity and rotations. What this means is that in the ensemble any
fluctuation is just as likely as its parity-reversed, or rotated counterpart.
The implication of this for the statistics of the CMB is that any expectation
value must be preserved if we rotate or invert the fields involved.
For the two-point statistics, this limits the non-zero correlations between
the observable multipoles to
\begin{eqnarray}
\langle T_{lm} T^*_{l'm'} \rangle = \delta_{ll'} \delta _{mm'} C_l^T \; ,
\quad &&
\langle E_{lm} E^*_{l'm'} \rangle = \delta_{ll'} \delta _{mm'} C_l^E \; ,
\nonumber \\
\langle B_{lm} B^*_{l'm'} \rangle = \delta_{ll'} \delta _{mm'} C_l^B \; ,
\quad &&
\langle T_{lm} E^*_{l'm'} \rangle = \delta_{ll'} \delta _{mm'} C_l^{TE} \; .
\label{adc:eq10}
\end{eqnarray}
For Gaussian fluctuations, such as those predicted by the simplest models
of inflation, these power spectra $C_l^T$, $C_l^E$, $C_l^B$ and $C_l^{TE}$
contain all of the statistical information in the CMB. There have been several
claims of violations of Gaussianity/rotational invariance in the one-year
WMAP temperature data, e.g.~\cite{adc:oliveira04,adc:vielva04,adc:copi04,adc:eriksen04,adc:hansen04}, but the issue of whether these
are of primordial origin, or rather artifacts of the instrument or local
Universe, is unresolved~\cite{adc:schwarz04}.
Example power spectra of the CMB observables for adiabatic density
perturbations in a flat, $\Lambda$CDM model are shown in Fig.~\ref{adc:fig2}.
The physics that gives rise to these spectra is reviewed briefly
in Sect.~\ref{adc:secphysics}.

\begin{figure}[t!]
\begin{center}
\parbox{0.49\textwidth}{
\includegraphics[angle=-90,width=0.45\textwidth]{adc_f4.ps}
}
\parbox{0.49\textwidth}{
\includegraphics[angle=0,width=0.45\textwidth]{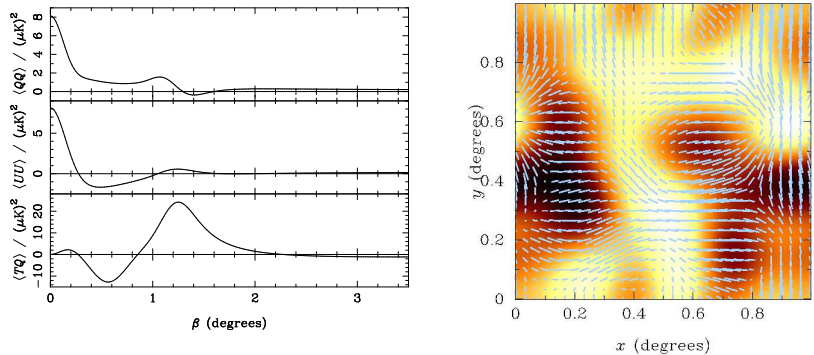}
}
\end{center}
\caption{Polarization correlation functions for adiabatic density
perturbations (left). The right-hand plot overlays a simulation
of the \emph{correlated part} of the polarization on the CMB temperature
anisotropies.
}
\label{adc:fig3} 
\end{figure}

It is also interesting to consider the two-point correlations between
the Stokes parameters (and also the temperature anisotropies) in real
space~\cite{adc:coulson94}.
For polarization, the correlation will be manifestly
rotationally-invariant only if we work with Stokes parameters defined
on a basis intrinsic to the two points, $\vnhat_1$ and $\vnhat_2$, under
consideration.
If we define the local $x$-direction by the geodesic
connecting the two points, and denote the polarization on these bases by
an overbar, the correlations are~\cite{adc:kamionkowski97,adc:ng99,adc:chon04}
\begin{eqnarray}
\langle \bar{Q}(\vnhat_1) \bar{Q}(\vnhat_2) \rangle &=&
\frac{1}{2} \sum_l \frac{2l+1}{4\pi}[C_l^E (d^l_{2\,2}
+ d^l_{2\,-2})(\beta) + C_l^B (d^l_{2\,2} - d^l_{2\,-2})(\beta)] \; ,
\nonumber \\
\langle \bar{U}(\vnhat_1) \bar{U}(\vnhat_2) \rangle &=&
\frac{1}{2} \sum_l \frac{2l+1}{4\pi}[C_l^E (d^l_{2\,2} 
- d^l_{2\,-2})(\beta) + C_l^B (d^l_{2\,2} + d^l_{2\,-2})(\beta)] \; ,
\nonumber \\
\langle T(\vnhat_1) \bar{Q}(\vnhat_2) \rangle &=&
\sum_l \frac{2l+1}{4\pi}C_l^{TE} d^l_{2\,0}(\beta) \; , \nonumber \\
\langle T(\vnhat_1) T(\vnhat_2) \rangle &=&
\sum_l \frac{2l+1}{4\pi}C_l^{T} d^l_{00}(\beta) \; ,
\label{adc:eq11}
\end{eqnarray}
where $\beta$ is the angle between the two points ($0 \leq \beta \leq \pi$),
and $d^l_{mn}(\beta)$ are the reduced Wigner functions. The correlations
$\langle \bar{Q} \bar{U} \rangle $ and $\langle T \bar{U} \rangle$
vanish due to parity and rotational invariance. Examples of the
non-zero polarization correlation functions are plotted in Fig.~\ref{adc:fig3},
along with a simulated Gaussian realisation of the temperature anisotropies
on a flat patch of the sky with the \emph{correlated part} of the polarization
overlaid. As first noted in~\cite{adc:coulson94}, the negative tail to the
temperature--polarization correlation function on large scales is generic,
arising from the infall of the photon--baryon plasma into potential wells
around recombination, and gives a tangential pattern of correlated polarization
around large-scale temperature hot spots. On smaller scales the
sign of the cross-correlation oscillates. For adiabatic perturbations
on scales smaller than $\sim 0.3^\circ$, the cross-correlation is positive and
this gives rise to the tangential polarization pattern around small-scale
cold spots that can be seen in Fig.~\ref{adc:fig3}.

\subsubsection{Cosmic Variance}
\label{adc:subseccosmic}

Since for Gaussian CMB fluctuations all cosmological information is
encoded in the various power spectra, the precision with which the
CMB can constrain cosmology is determined, in part, by the accuracy
to which these spectra can be determined.
In the absence of instrumental noise, and with full sky coverage,
cosmic variance provides a fundamental limit to this accuracy.
Cosmic variance arises from having a single realisation
of a non-ergodic process (the sphere is compact) from which to estimate
the power spectra. In this case, the obvious estimators of the power spectra,
e.g.\
\begin{equation}
\hat{C}_l^E = \frac{1}{2l+1} \sum_m |E_{lm}|^2 \; ,
\label{adc:eq12}
\end{equation}
are optimal in that they have the minimum variance of all unbiased estimators.
At each $l$ we have only $2l+1$ independent real numbers from which to estimate
a variance, so the cosmic variance in our estimator $\hat{C}_l^E$ for Gaussian
fields is
\begin{equation}
\mathrm{var}(\hat{C}_l^E ) = \frac{2}{2l+1} (C_l^E)^2 \; .
\label{adc:eq13}
\end{equation}
Equivalent results apply to the two other auto-spectra $C_l^T$ and $C_l^B$.
For the cross-spectrum, the cosmic variance is only slightly more
complicated~\cite{adc:kamionkowski97,adc:zaldarriaga97stats}:
\begin{equation}
\mathrm{var}(\hat{C}_l^{TE} ) = \frac{1}{2l+1} \left[(C_l^{TE})^2
+ C_l^T C_l^E \right] \; .
\label{adc:eq14}
\end{equation}
The second term arises from chance correlations in the single realisation
that we have available, and would fundamentally limit the accuracy of
searches for parity violations via e.g.\ the cross-correlation between
the temperature anisotropies and magnetic polarization.
Due to the presence of correlations between the temperature and the
electric polarization, estimates of $C_l^T$ and $C_l^E$ have non-vanishing
covariance, and each is also correlated with the estimator for $C_l^{TE}$:
\begin{equation}
\mathrm{cov}(\hat{C}_l^T,\hat{C}_l^E) = \frac{2}{2l+1} (C_l^{TE})^2 \; ,
\qquad \mathrm{cov}(\hat{C}_l^{T/E},\hat{C}_l^{TE}) = \frac{2}{2l+1}
C_l^{T/E} C_l^{TE} \; .
\label{adc:eq15}
\end{equation}

Of course, in the realistic case of non-zero instrument noise and
limited sky coverage the precision of power spectrum estimates falls
short of the cosmic variance limit. The effect of only observing
a fraction $f_{\mathrm{sky}}$ of the sky is not only to increase the
variance~\cite{adc:scott94}, but also to correlate estimates across
$\Delta l$ given roughly by the inverse linear size of the retained portion
of the sky~\cite{adc:hobson96,adc:tegmark96}.
The increase in variance can be crudely accounted for 
by reducing the number of independent modes to
$(2l+1)f_{\mathrm{sky}}$. For polarization, this mode-counting
argument ignores the loss of ambiguous modes that cannot be properly
classified as electric or magnetic with partial sky coverage
(see Sect.~\ref{adc:secanalysis}). This effect is minimised for surveys
covering large connected regions of the sky.

\section{Physics of CMB Polarization}
\label{adc:secphysics}

When the mean free path was small compared to the spatial scale (i.e.\
wavelength for a plane wave) of the perturbations, Thomson scattering
kept the CMB radiation isotropic in the rest frame of the electron--baryon
plasma. As protons and electrons started to recombine to form neutral
hydrogen, the mean free path increased and anisotropies started to
develop. Subsequent Thomson scattering of the quadrupole component
of the anisotropy generated linear polarization; this was then preserved
for the free-streaming CMB until the Universe was reionized by the ionizing
radiation from the first non-linear structures.
The photon visibility function --
the probability that a photon last scattered around recombination as
a function of cosmic time -- peaks around $370\;\mathrm{kyr}$
after the big bang and
has a width $\sim 115\; \mathrm{kyr}$~\cite{adc:spergel03}.
Since polarization is
only generated by scattering it provides a very clean probe of conditions
around the time of recombination and reionization.

Consider unpolarized radiation with a temperature distribution
$\Theta(\ve)$, where $\ve$ is the radiation propagation direction.
If the radiation field possesses a quadrupole anisotropy $\Theta_2(\ve)$,
Thomson scattering generates linear polarization with a polarization
tensor~\cite{adc:hu97angmom,adc:challinor00}
\begin{equation}
\clp_{ab}(\ve) = - \frac{1}{20} \D\tau \nabla_{\langle a} \nabla_{b\rangle}
\Theta_2 \; ,
\label{adc:eq16}
\end{equation}
where $\D\tau$ is the differential optical depth, and the covariant derivatives
are in the surface $\ve^2 = 1$. Polarization is only generated from
the quaudrupole component of the incident radiation, and can be seen to
be purely electric quadrupole in character. If the incident radiation is
polarized, there is an additional contribution to the polarization produced
due to the electric quadrupole of the incident polarization ($\Theta_2$
should be replaced by
$\Theta_2 - 12 P_{E,2}$~\cite{adc:hu97angmom,adc:challinor00}).
For the CMB, the linear
polarization we observe (at $\vx=0$) along a line of
sight $\vnhat$ is that produced
at last scattering along the radiation direction $\ve = -\vnhat$, but
at a direction-dependent comoving position $\vx = \Delta \eta \vnhat$. (We are
assuming the Universe is flat for simplicity, and $\Delta \eta$ is the
conformal time between last scattering and the present.) The character
of the \emph{observed} polarization thus depends on the spatial distribution of
the polarization at last scattering which, in turn, depends on the type
of perturbations (e.g.\ density or gravitational waves) we consider.

\subsection{Density Perturbations}

\begin{figure}[t!]
\begin{center}
\includegraphics[angle=0,width=0.35\textwidth]{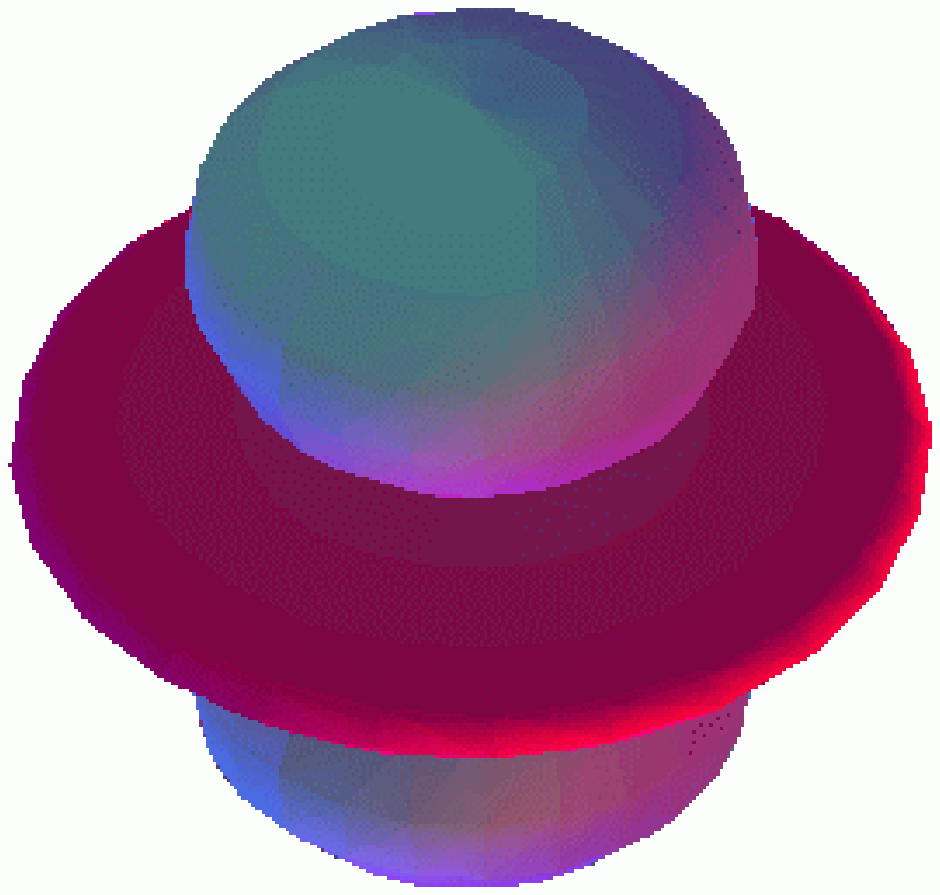}
\quad
\includegraphics[angle=0,width=0.35\textwidth]{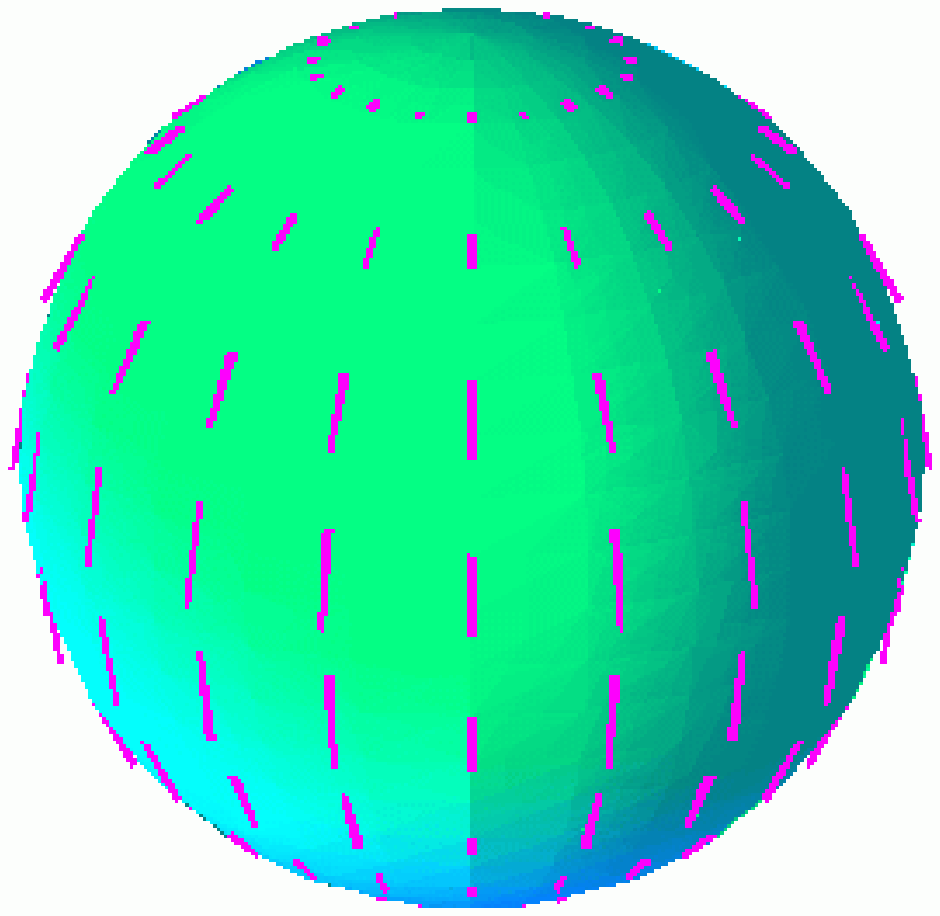}
\end{center}
\vspace{-2\baselineskip}
\caption{Generation of linear polarization by Thomson scattering for a
single Fourier mode of the density perturbations,
following the treatment of~\cite{adc:hu97primer}.
The temperature anisotropies
are azimuthally symmetric about the wavevector; the quadrupole part is shown
on the left with cold lobes along the $\pm z$-directions. The linear
polarization that is produced is then a pure $l=2$, $m=0$ electric polarization
pattern locally (right). To get the observed polarization after free
streaming from last scattering, the local polarization is modulated
by the plane wave $\exp(i\vk\cdot \vnhat \Delta \eta)$.
}
\label{adc:fig5} 
\end{figure}

If we consider a single plane-wave density perturbation of comoving wavenumber
$k$, the peculiar velocity $\vvb$ of the electron--baryon plasma around
recombination is in the
direction of the wavevector $\vkhat$. Over the mean free time since the
previous scattering, a temperature quadrupole
\begin{equation}
\Theta_2(\ve) \sim - \ell_p T_{\mathrm{CMB}}
\nabla_{\langle i} v_{\mathrm{b},j \rangle} e^i e^j
\propto T_{\mathrm{CMB}} \ell_p k P_2(\vkhat \cdot \ve)
\end{equation}
is produced to leading order in
$\ell_p k$ where $\ell_p$ is the mean free path. The quadrupole is azimuthally
symmetric about the wavevector, and Thomson scatters to generate linear
polarization parallel to the projection of $\vkhat$ on the sky (see
Fig.~\ref{adc:fig5}). The polarization tensor at last scattering has a spatial
dependence $\exp(\I \vk \cdot \vx)$ appropriate to the plane wave, and this
leads to a further modulation of the polarization observed today by
$\exp(\I \Delta \eta \vk \cdot \vnhat)$. This modulation
transfers polarization from the $l=2$ to higher multipoles (with most
appreciable power appearing at $l = k \Delta \eta$), but preserves the
electric character of the polarization~\cite{adc:seljak97,adc:kamionkowski97}.
To see that this is
reasonable, note that the modulated polarization field has its
polarization direction either parallel or perpendicular to the direction in
which the polarization amplitude is changing~\cite{adc:hu97primer}.
Comparison with
Fig.~\ref{adc:fig1} shows that such a pattern is pure electric polarization.
This important observation, that linear density perturbations do not
produce magnetic polarization, can also be understood from the fact that
the spatial distribution of $\nabla_{\langle i}
v_{\mathrm{b},j\rangle}$ is curl-free~\cite{adc:challinor00}.

An example of the polarization power spectra produced by adiabatic density
perturbations is plotted in Fig.~\ref{adc:fig2}. They were computed
with the Boltzmann code CAMB~\cite{adc:lewis00}.
The $E$-mode power peaks around
$l \sim 1000$, corresponding to the angle subtended by the width of the
visibility function at recombination. On larger scales
the polarization probes the electron--baryon velocity at last
scattering, as described above. Acoustic oscillations of the plasma
prior to last scattering imprint an oscillatory structure on the angular
power spectra of the CMB observables. Fourier modes of the photon density
that are caught at the
extrema of their oscillation at last scattering produce the peaks in the
temperature-anisotropy power spectrum, but troughs in the polarization since
$\vvb$ then vanishes for such modes. Modes caught at either the extrema or
the midpoint of their oscillation give zeroes in the temperature--polarization
cross-correlation. Large-angle polarization from the last scattering
surface is very small: the optimal wavelength for generating power at
scale $l$ is $l / \Delta \eta$, but this is too small to allow
significant generation of quadrupole temperature anisotropies over a mean
free path before last scattering. (Large-angle
polarization is further suppressed since $\vvb$ is small on super-Hubble
scales.) The increase in polarization on large scales in Fig.~\ref{adc:fig2}
is due to reionization~\cite{adc:zaldarriaga97reion}. Re-scattering of the
temperature quadrupole generates linear polarization peaking at multipoles
roughly
twice the ratio of the conformal look-back time to reionization to
the conformal time elapsed between recombination and reionization. The
amplitude of the polarization produced at reionization is proportional
to the number of photons that scatter (i.e.\ the optical depth to reionization,
taken to be $\tau = 0.15$ in Fig.~\ref{adc:fig2}).

Although linear density perturbations do not produce $B$-mode polarization,
this is produced by a number of second-order effects. The most notable is
weak gravitational lensing by large-scale
structure~\cite{adc:zaldarriaga98lens}, which
effectively re-maps the polarization on the sky according to the lensing
deflection field. This can distort a curl-free pattern into one with
curl and so generate magnetic polarization. The power spectrum of this
effect is also included in Fig.~\ref{adc:fig2}. On large scales the
spectrum is almost white with $C_l^B \sim (1.3 \, \mathrm{nK})^2$.
The lens-induced $B$-mode spectrum peaks at $l \sim 1000$ corresponding to
the peak in $C_l^E$.

\subsection{Gravitational Waves}

\begin{figure}[t!]
\begin{center}
\includegraphics[angle=0,width=0.35\textwidth]{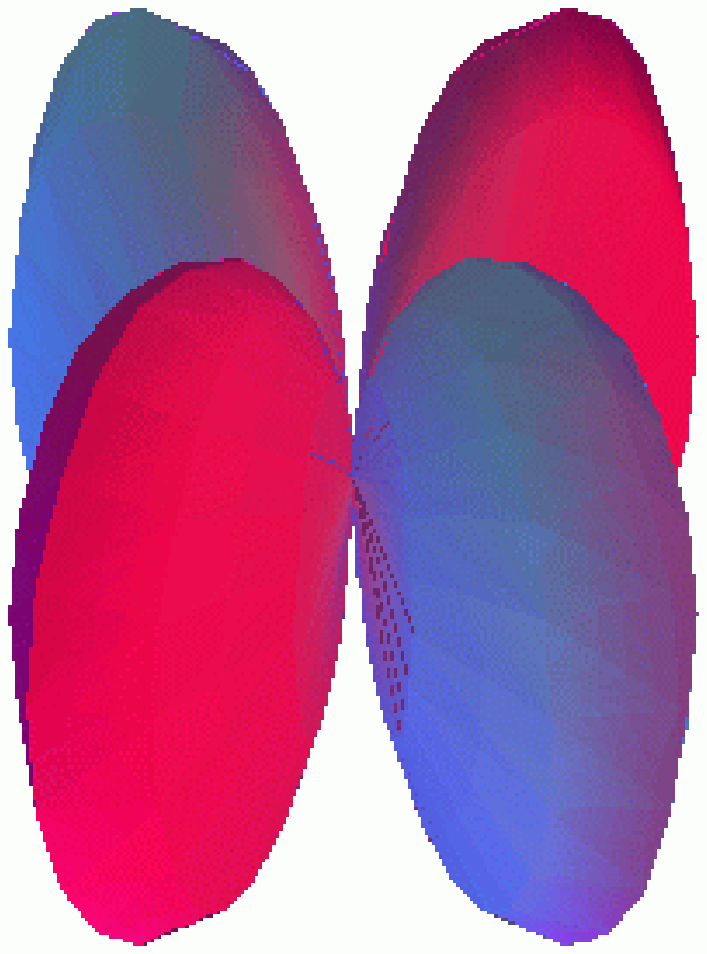}
\quad
\includegraphics[angle=0,width=0.35\textwidth]{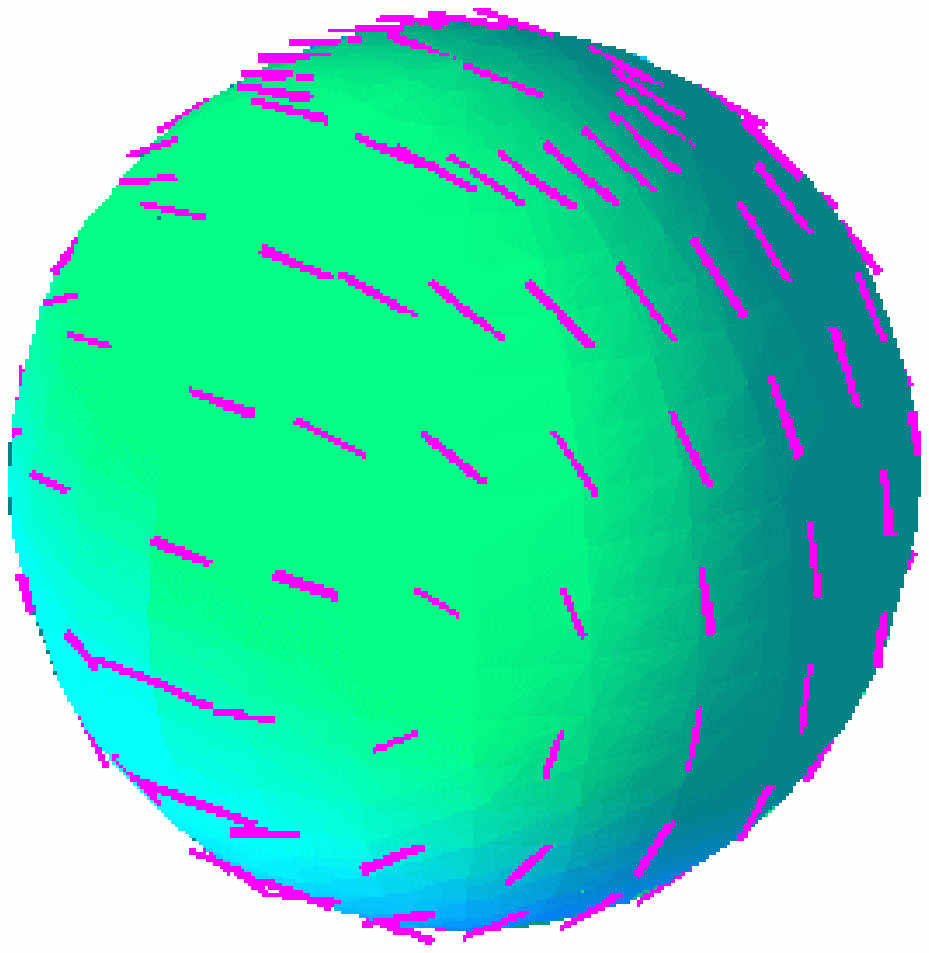}
\end{center}
\vspace{-2\baselineskip}
\caption{As in Fig.~\ref{adc:fig5} but for a Fourier mode of the
gravitational wave perturbations. The quadrupole part of the temperature
anisotropy now has $m=\pm 2$ modes (left); the hot lobes correspond to
the directions in which space is locally being contracted by the wave.
The linear polarization produced by Thomson scattering (right) is again
electric and quadrupole in character at the last scattering surface.
Unlike density perturbations, magnetic polarization is now generated by free
streaming since the polarization field modulated by the plane wave has
a non-zero curl.
}
\label{adc:fig6} 
\end{figure}

A stochastic background of gravitational waves, as predicted in all
inflationary models, can also influence the CMB. Locally,
a gravitational wave causes a quadrupole anisotropy in the
expansion of space, so photons arriving at a scatterer from different
lines of sight will suffer different redshifts since their previous
scattering event. The longest wavelength gravitational waves
are geometrically most efficient at generating a quadrupole moment.
For wavelengths long compared to the mean free path, the quadrupole moment
$\Theta_2(\ve) \propto - \ell_p T_{\mathrm{CMB}} \dot{h}_{ij} e^i e^j /2$,
where $h_{ij}$ is
the transverse, trace-free perturbation to the Robertson--Walker metric,
\begin{equation}
\D s^2 = a^2(\eta) [\D \eta^2 - (\delta_{ij} + h_{ij}) \D x^i \D x^j] \; ,
\label{adc:eq17}
\end{equation}
and an overdot denotes differentiation with respect to proper time. 
However, the shear of the gravitational wave $\dot{h}_{ij}$ is zero for
wavelengths well outside the horizon, and undergoes damped oscillation inside
the horizon.
It follows that the temperature quadrupole at last scattering is
dominated by modes with wavelength of the order of the
horizon size, and this sets the
characteristic scale ($l \sim 100$) on which the polarization power spectra
peak (see Fig.~\ref{adc:fig2}). Thomson scattering of the temperature
quadrupole from gravitational waves generates a polarization tensor
at last scattering $\clp_{ab} \propto [\dot{h}_{ab}]^{\mathrm{TT}}$,
i.e.\ proportional to the trace-free projection of the shear onto the
sphere $\ve^2 =1$. For a single plane wave with wavevector along the $z$-axis,
the quadrupole moment and local polarization from the plane wave have only
$m=\pm 2$ modes (see Fig.~\ref{adc:fig6} and~\cite{adc:hu97angmom}
for further details).
The observed polarization from a plane wave is further modulated by
$\exp(\I \Delta \eta \vk \cdot \vnhat)$ in free-streaming from last
scattering. Modulating the polarization in Fig.~\ref{adc:fig6} gives
a polarization pattern that locally looks electric along the lobes of the
temperature quadrupole, and magnetic in between~\cite{adc:hu97primer}.
It follows that
both electric and magnetic polarization are generated by gravitational
waves~\cite{adc:seljak97,adc:kamionkowski97},
and with roughly equal powers~\cite{adc:hu97angmom}.
The generation of $B$-mode
polarization from gravitational waves can also be understood as a consequence
of the non-vanishing (spatial) curl of the shear $\dot{h}_{ij}$ at last
scattering~\cite{adc:challinor00}.

The power spectra of the CMB observables from a scale-invariant background of
gravitational waves is shown in Fig.~\ref{adc:fig2}. As for
density perturbations, large-angle polarization is generated by reionization.
The amplitude of any gravitational wave background is currently only
weakly constrained by CMB observations. A recent analysis, combining
CMB observations and data from the Sloan Digital Sky Survey (SDSS),
constrained the tensor-to-scalar ratio $r$ -- a measure of the ratio of the
primordial power in gravitational waves to curvature (density)
perturbations -- to $r< 0.36$ at 95\% confidence in a seven-parameter flat
model.
This is similar to the value adopted in Fig.~\ref{adc:fig2}, and shows
that the gravitational wave contribution to the temperature anisotropies
and electric polarization is sub-dominant.

\section{Current Status of CMB Polarization Measurements}
\label{adc:secmeasure}

\begin{figure}[t!]
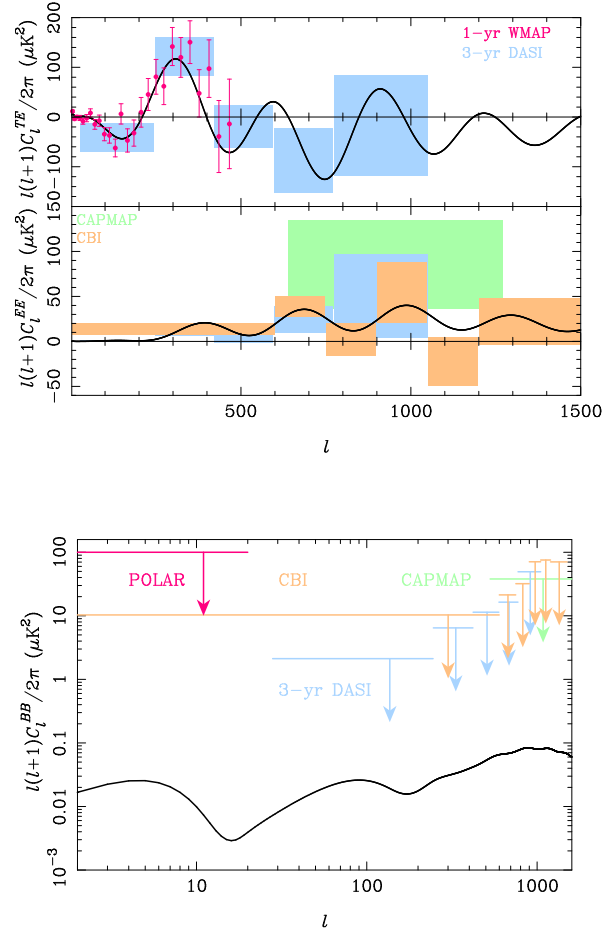

\begin{center}
\includegraphics[width=6cm,angle=-90]{adc_f7a.ps}
\end{center}
\begin{center}
\hspace{0.7cm}
\includegraphics[width=6cm,angle=-90]{adc_f7b.ps}
\end{center}
\caption{Current polarization detections and best upper limits
(as of January 2005). The top plot shows $C_l^{TE}$ measurements (top)
and $C_l^E$ (bottom). The points with 1-$\sigma$ errors are from the first
one-year data release from WMAP.
The error boxes are the flat band-power results from
DASI, CBI and CAPMAP data,
centred on the maximum-likelihood
band power and spanning the 68-per cent intervals. The solid lines are the
predicted power from the best-fit model to all the WMAP data. The bottom
plot shows the upper limits on $C_l^B$ from DASI, CBI, CAPMAP and
POLAR. The solid line assumes a tensor-to-scalar ratio
$r=0.38$.
}
\label{adc:fig7} 
\end{figure}

The first detection of polarization of the CMB was announced
in September 2002~\cite{adc:kovac02}. The measurements were made with DASI,
a compact interferometric array operating at 30\,GHz, deployed at the
South Pole. The DASI team have now analysed three years of data; they
report a detection of $E$-mode polarization at high significance
($6.3\sigma$) and at a level perfectly consistent with predictions
based on temperature-anisotropy data~\cite{adc:leitch04}. The only
other reported direct detections to date are from the Cosmic Anisotropy
Polarization MAPper (CAPMAP;~\cite{adc:barkats05}),
a heterodyne correlation polarimeter
operating at $90\, \mathrm{GHz}$ from New Jersey, and from the
Cosmic Background Imager (CBI;~\cite{adc:readhead04}), an
interferometer operating
in the range 26--$36\, \mathrm{GHz}$ from Chile. The results on $C_l^E$
from maximum-likelihood power spectrum analyses by the
DASI, CBI and CAPMAP teams are shown in Fig.~\ref{adc:fig7}, along with
the prediction based on the best-fit to all the first-year WMAP data.
Although the data are very noisy and broad-band (due to the limited sky
coverage), the agreement with the prediction is striking. In addition,
the $TE$
cross-correlation has been measured with good precision by
WMAP~\cite{adc:kogut03},
and has also been detected by DASI (see Fig.~\ref{adc:fig7}).

The $B$-mode polarization of the CMB is currently subject only to
weak upper limits.
The best limit on large scales, $l\sim 10$, is from
the POLAR experiment~\cite{adc:keating01}, while the best limits on smaller
scales are from DASI~\cite{adc:leitch04}. These (95\%) upper limits
are shown in Fig.~\ref{adc:fig7}, along with the predicted $B$-mode
spectrum in a model with tensor-to-scalar ratio $r=0.38$, close to
the current 95\% limit from temperature anisotropies and
SDSS~\cite{adc:seljak04}.
Current direct constraints on the amplitude of gravitational waves
from $B$-mode data are clearly much weaker than this.
Furthermore, upper limits on the $B$-mode power on small scales are at
least two orders of magnitude above the level expected from gravitational
lensing.

\section{What do we learn from CMB polarization?}
\label{adc:secmotivation}

Electric polarization primarily traces the velocity of the electron--baryon
plasma at recombination due to density perturbations. It thus provides
complementary information to the temperature anisotropies which are sourced
mainly by the photon density variations at last scattering on sub-degree
scales. Polarization observations can thus be expected to bring significant
improvements on some of the `acoustic' cosmological
parameters~\cite{adc:zaldarriaga97inf},
such as the physical density of baryons and cold
dark matter, although current polarization data is not yet of sufficient
quality relative to the temperature data to see this
effect~\cite{adc:readhead04}. Furthermore, the \emph{relative} phase of the
acoustic peaks in the temperature and electric-polarization power
spectra (and the cross-correlation) can be used for a largely
model-independent test of the physics of acoustic oscillations.
For both adiabatic and isocurvature perturbations the acoustic
peaks in $C_l^T$ and $C_l^E$ should be in antiphase. (Admixtures of
these perturbations can, however, partially destroy this coherence.)
A recent analysis by the CBI team~\cite{adc:readhead04} found the
phase of the oscillations in the $E$-mode spectrum to be fully
consistent with expectations for adiabatic models.
The spectrum $C_l^{TE}$ from the one-year WMAP data~\cite{adc:kogut03} has also
been used to improve constraints on the amplitude of any isocurvature
fluctuations, e.g.~\cite{adc:peiris03,adc:gordon03,adc:bucher04}.

Polarization on scales of tens of degrees ($l < 10$) can be used to probe
reionization. Significantly, it breaks the degeneracy between
the amplitude of density perturbations, $A_\mathrm{s}$, and the optical
depth to reionization, $\tau$, that leaves only $A_\mathrm{s} e^{-2\tau}$
well-determined from the temperature anisotropies.
(The power spectra of the CMB observables are reduced by
$e^{-2\tau}$ on scales corresponding to Fourier modes that are sub-horizon at
reionization.) The one-year
WMAP measurement of $C_l^{TE}$ shows a significant excess power on large
scales that is attributed to early reionization ($11 < z_\mathrm{re} < 30$)
giving a significant optical depth $\tau = 0.17 \pm 0.04$
(68\% interval)~\cite{adc:kogut03}.
The amplitude of curvature (density) perturbations (at $0.05\, \mathrm{Mpc}$)
is then constrained to
$A_\mathrm{s} = (2.7 \pm 0.3) \times 10^{-9}$ from the CMB
alone~\cite{adc:spergel03}.
Looking further ahead, the fine details of the large-angle polarization power
can, in principle, distinguish different ionization histories with the
same optical depth, although this is hampered by the large cosmic variance
at low $l$~\cite{adc:holder03,adc:hu03reion}.

Perhaps the greatest ambition for CMB polarization observations is the
detection of the $B$-mode signal from gravitational waves. Although
gravitational waves do generate temperature anisotropies and electric
polarization, they
do so only on large scales where the cosmic variance of the dominant
density perturbations is large. Indeed, a perfect temperature-anisotropy
experiment can only detect $r>0.07$, even assuming all other parameters are
known. Supplementing the temperature anisotropies with electric polarization
makes only a modest further improvement: $r>0.02$. Although the current upper
limit on $r$ from $B$-mode polarization is much weaker than that inferred
from the CMB temperature anisotropies (see Fig.~\ref{adc:fig7}), the former
route ultimately has the potential to probe much lower values as thermal
noise levels improve. Assuming that astrophysical foregrounds can be
removed adequately, the ultimate limit to a detection of the gravitational
wave signal is set by the $B$-modes generated by gravitational
lensing~\cite{adc:knox02,adc:kesden02,adc:seljak04lens}.

A detection of a stochastic background of gravitational waves would
have significant implications for fundamental physics. In inflation models,
the power produced in gravitational waves as a function of wavelength is a
direct measure of the expansion rate of the Universe at the time that
wavelength was stretched beyond the horizon. In slow-roll inflation in a
potential $V$, the power in gravitational waves thus constrains the
`energy scale of inflation', $V^{1/4}$, via
\begin{equation}
V^{1/4} = 3.33 \times 10^{16} r^{1/4} (A_\mathrm{s}/2.3\times 10^{-9})^{1/4}\,
\mathrm{GeV} \; .
\label{adc:eq18}
\end{equation}
The current 95\% limit $r < 0.36$~\cite{adc:seljak04} constrains $V^{1/4} < 2.3
\times 10^{16} \, \mathrm{GeV}$. By way of contrast,
the recently-introduced cyclic model~\cite{adc:steinhardt02} would produce
negligible gravitational waves as the fluctuations are produced during
a phase of very slow contraction~\cite{adc:boyle04}.
A detection of cosmological
gravitational waves would effectively rule out such models. In addition
to constraining the energy scale of inflation, measuring $r$ is important
for constraining the \emph{shape} of the inflaton potential. For the simplest
models of slow-roll inflation, the observables $r$ and $n_\mathrm{s}$ (the
spectral index of curvature perturbations) are both required to specify
the first two derivatives of the inflaton potential. Without the tight
limits on $r$ from future $B$-mode searches, constraints on inflation
models will be largely degenerate along the direction of constant
$n_\mathrm{s}$. The most recent constraints on $r$ and $n_\mathrm{s}$
from the CMB and galaxy surveys
already rule out some popular models such as
$V \propto \phi^4$~\cite{adc:seljak04}.

There are several other potential sources of $B$-mode polarization that could
dominate over any signal of primordial gravitational waves from inflation.
A sub-dominant background
of cosmic strings -- currently enjoying a resurgence of interest following
the realisation that they are naturally produced in some models of brane
inflation~\cite{adc:jones02} -- would provide a seed for gravitational waves
(and vortical, or vector, modes) on the last scattering surface.
Both give rise to
$B$-mode polarization, and the resulting power spectrum is expected to
peak on sub-degree scales~\cite{adc:seljak97defect}. It has been
argued that a statistically-isotropic, but highly
non-Gaussian, pattern of $B$-mode polarization with power peaking on small
scales would be a smoking-gun signature for a background of cosmic
strings~\cite{adc:pogosian03}.
In addition, primordial magnetic fields have several interesting
effects on the CMB and its polarization, although the theory is still
rather under-developed. The anisotropic stress of a stochastic field
also excites gravitational waves and vector modes (e.g.~\cite{adc:lewis04})
leading to (non-Gaussian) $B$-mode polarization. Faraday
rotation by either
large-scale coherent~\cite{adc:kosowsky96,adc:scannapieco97} or stochastic
fields~\cite{adc:kosowsky04}
can also rotate electric polarization into magnetic. The hope is that
these effects could be separated on the basis of their non-Gaussian
statistical properties,
and, in the case of Faraday rotation, by its frequency dependence.

Finally, we note that weak gravitational lensing effects on the CMB are
a valuable source of additional cosmological information,
most notably on the properties of dark energy and neutrino
masses~\cite{adc:hu02synergy}. As we describe further in
Sect.~\ref{adc:seclensing}, CMB polarization is particularly useful here
since intrinsically it has more power on small scales than the temperature
and so is more sensitive to the small-scale power in the lensing deflection.

\section{Polarization data analysis}
\label{adc:secanalysis}

The CMB polarization signal is small. Current measurements are consistent with
the expected r.m.s.\ signal $\sim 6 \, \mu\mathrm{K}$, which should be
compared with the $\sim 120 \, \mu\mathrm{K}$ r.m.s.\ of the temperature
anisotropies.
The low signal level requires high thermal sensitivity and demands
exquisite control of systematic effects.
The impact on the final science products of any systematics that cannot be
designed out of the instrument must be
carefully assessed (usually with a large number of simulations), and
any that are found to be significant must be accounted for in the data
analysis. As noted in the Introduction, this inevitably makes the analysis
of polarization data more instrument-specific than for temperature-anisotropy
experiments. Despite this, there are a number of common steps in the
analysis pipeline of any polarization experiment: map-making from time-series
data; astrophysical foreground removal; $E$-$B$ mode separation; power
spectrum estimation; and parameter estimation. (Separation of $E$ and $B$ modes
can also be performed statistically during the estimation of power spectra, and
this has been the case for all analyses of real data to date.) In the
following subsections we briefly summarise most of these key steps, but focus
on $E$-$B$ separation and power spectrum estimation in particular.
While the other steps are, of course, crucially important, the analysis methods
employed there are relatively straightforward generalisations of those
developed for the temperature anisotropies, for which excellent reviews
already exist (see e.g.~\cite{adc:bond01}). 

\subsection{Modelling the polarimeter response}

We begin by looking at how to model the
response of a polarimeter, i.e.\ the relation between the observed data
in the time domain to the fields on the sky.
The most general (instantaneous) linear response of a polarimeter to the
polarized sky brightness at each frequency is of the form
\begin{equation}
s \propto \int (\tilde{I} I + \tilde{Q} Q + \tilde{U} U - \tilde{V} V) \,
\D \vnhat \;
\label{adc:eq19}
\end{equation}
Here, $I(\vnhat,\nu)$, $Q(\vnhat,\nu)$, $U(\vnhat,\nu)$ and $V(\vnhat,\nu)$
are the Stokes brightness parameters at frequency $\nu$, and the
equivalent quantities with tildes refer to the instrument response.
The latter parameters are determined by the instrument optics and
receiver. They inherit a time-dependence from the scanning of the
instrument on the sky -- $\tilde{I}(\vnhat,\nu)$ and $\tilde{V}(\vnhat,\nu)$
rotate with the instrument as scalar functions and $\tilde{Q}(\vnhat,\nu)$ and
$\tilde{U}(\vnhat,\nu)$ as components of a spin-2 function -- but also from
any modulation scheme employed to ameliorate the effects of low-frequency
noise in the instrument. The signal contribution to the data is
given by further integrating~(\ref{adc:eq19}) over the spectral response of the
instrument.

A useful idealisation is an instrument with ideal optics and receiver, and
a narrow spectral bandpass. For a dual-polarization system, such
as a pair of polarization-sensitive bolometers on the high-frequency
instrument of the Planck satellite, the beam patterns on the sky
for the two polarization modes should have field directions that are
orthogonal and constant, and the amplitude of the response should be
azimuthally symmetric. In the case of the bolometric system, the ideal
instrument response is then fully determined by $\tilde{I}(\vnhat)$,
which is a function of $\theta$ only. The response to circular polarization
vanishes and
$\tilde{Q} \pm \I \tilde{U} = - \tilde{I}(\theta) \E^{\pm 2 i\phi}$ for
the $y$-polarization, and minus this for the $x$-polarization. If we now
rotate the instrument so the optic axis is along the direction $\vnhat$,
and the local $y$-polarization direction is at an angle $\psi$ to the
$\phi$-direction there, the signal received in the $y$-polarization
simplifies to~\cite{adc:challinor00convolve}
\begin{equation}
\int (\tilde{I} I + \tilde{Q} Q + \tilde{U} U - \tilde{V} V) \,
\D \vnhat = I_{\mathrm{s}}(\vnhat) - Q_{\mathrm{s}}(\vnhat) \cos 2 \psi
+ U_{\mathrm{s}}(\vnhat) \sin 2 \psi \; ,  
\label{adc:eq20}
\end{equation}
where the beam-smoothed fields are
\begin{eqnarray}
I_{\mathrm{s}}(\vnhat) &=& \sum_{lm} W_l T_{lm} Y_{lm}(\vnhat) \; ,
\label{adc:eq21} \\
(Q_{\mathrm{s}} \pm \I U_{\mathrm{s}})(\vnhat) &=& \sum_{lm}
{}_2 W_l (E\mp \I B)_{lm} {}_{\mp 2} Y_{lm}(\vnhat) \; .
\label{adc:eq22}
\end{eqnarray}
Here, the $W_l$ is proportional to the $m=0$ multipoles of
$\tilde{I}(\theta)$ and ${}_2 W_l$ to the $m=2$ electric multipoles
of $\tilde{Q} \pm \I \tilde{U}$. For a Gaussian beam of dispersion
$\sigma \ll 1$, we have beam functions
\begin{equation}
W_l \approx e^{-l(l+1) \sigma^2/2} \quad \mbox{and} \quad
{}_2 W_l \approx e^{-[l(l+1)-4] \sigma^2/2} \; .
\label{adc:eq23}
\end{equation}
For $l \gg 1$ the polarized and total intensity beam functions
are very nearly equal. The form of~(\ref{adc:eq20}) is equivalent to a point
sampling of the smoothed fields along the optic axis with a polarizer
in orientation $\psi$. The signal in the $x$-polarization is got by
replacing $\psi$ by $\psi - \pi/2$. 
The generalisation of~(\ref{adc:eq20})
to other ideal polarimeters is straightforward: in each case the
signal remains a linear combination of the beam-smoothed fields.

\subsection{Polarized map making}

The aim of map-making is two-fold: (i) to produce pixelised images of the
temperature and polarization fields on the sky from the observed data;
and (ii) to compress the time-stream data to a more manageable size, while
preserving as much of the cosmological information present as possible.
We shall only consider the ideal instrument model described above;
for a case study of some of the real-world problems
that must be solved in the map-making process for total-intensity data
see~\cite{adc:stompor02}. For the ideal model, the signal contribution to the
data is linear in the beam-smoothed fields. Instrument noise is assumed
additive, in which case our simple model for the data from the $i$th
detector at time $t$ becomes
\begin{equation}
d^i_t = A^i_{tp}(I_{\mathrm{s},p} - Q_{\mathrm{s},p} \cos 2\psi^i_t +
U_{\mathrm{s},p} \sin 2\psi^i_t) + n_t^i \; .
\label{adc:eq24}
\end{equation}
Here, $A^i_{tp}$ is the detector pointing matrix and is unity if
the beam centre of detector $i$ points towards pixel $p$ at time $t$, and
is zero otherwise. The $i$th detector at time $t$ is sensitive to
polarization along a direction at angle $\psi^i_t$ to the $\phi$-direction
on the sky, and has instrument noise $n^i_t$.
We assume that all detectors have the same narrow spectral bandpass,
and the same angular resolution, so that we can combine the signals
of all detectors into a single vector $\vd$ and the instrument
noise into $\vn$.
We attempt to reconstruct the pixelised beam-smoothed fields, which we
represent by the vector $\vs$, by a regularised inversion of
$\vd = \mA \vs + \vn$, where the $(it)p$ element of $\mA$ is $A^i_{tp}$.

If the noise is Gaussian, with known
covariance $\langle \vn \vn^T \rangle = \mN$, the maximum-likelihood
solution,
\begin{equation}
\hat{\vs} = (\mA^T \mN^{-1} \mA )^{-1} \mA^T \mN^{-1} \vd \; ,
\label{adc:eq25}
\end{equation}
is optimal in the sense of being the minimum-variance unbiased estimate.
The covariance of the errors is $\langle (\hat{\vs} - \vs) (\hat{\vs}-\vs)^T
\rangle = (\mA^T \mN^{-1} \mA)^{-1}$. If the CMB fields are Gaussian,
the maximum-likelihood map contains all of the cosmological information
present in the time-stream data. In the language of statistics,
the maximum-likelihood map is a sufficient statistic for parameter
estimation, i.e.\ the data $\vd$ only enters the posterior probability
of the chosen (cosmological) parameters $\vec{\alpha}$ given the data,
$\mathrm{Pr}(\vec{\alpha} | \vd)$,
through $\hat{\vs}$. Forming the maximum-likelihood map can thus also be viewed
as lossless compression of the time-stream data. The question of the
feasibility of computing the maximum-likelihood map is discussed in discussed
in detail elsewhere in this volume by Borrill. Here we make just a few
general remarks.
\begin{itemize}
\item Any matrix can be used in place of the noise covariance matrix
$\mN$ in~(\ref{adc:eq25}) and the solution is still unbiased. However,
if the matrix used does not properly capture the noise correlations described
by off-diagonal elements of $\mN$, the resulting maps will have correlations
along the scan directions (`stripes').
\item Brute-force evaluation of $\hat{\vs}$ is prohibitive for large
datasets so approximate methods must be used. For stationary noise,
$\mN$ is circulant and can be applied efficiently in the Fourier domain.
Furthermore, direct inversion of $\mA ^T \mN^{-1} \mA$ can be avoided
by solving $\mA^T \mN^{-1} \mA \hat{\vs} = \mA^T \mN^{-1} \vd$ with
iterative techniques. 
\item For stationary noise, the power spectrum (the diagonal elements
of $\mN$ in Fourier space) can be estimated from the data if it is not
known a priori. Errors in the noise estimation increase the errors in the
map.
\item Maximum-likelihood map-making is easily generalised to include
template fitting and marginalisation. For example, the unknown amplitude
of some additive systematic effect whose `shape' is known can be marginalised
over during the map-making stage.
\item Fast, instrument-specific alternatives to maximum-likelihood map-making
have been proposed, e.g.\ the `destriping' technique for
Planck~\cite{adc:delabrouille98}.
\end{itemize}

\subsection{Astrophysical foreground removal}

The dominant diffuse polarized foregrounds in the frequency range
$20$--$400\, \mathrm{GHz}$ are expected to be Galactic synchrotron and
vibrational (thermal) dust emission. Synchrotron radiation in a uniform
magnetic field is linearly polarized. The polarization fraction depends on
the energy spectrum of the electrons, but can plausibly be as high as
70\%. Thermal dust emission will also be linearly polarized if
non-spherical grains are aligned in a magnetic field. Any variation of the
Galactic magnetic field along the line of sight will tend to depolarize
the radiation reducing the polarization fraction. This depolarization
tends to cancel the increased emission along the plane making the
polarization dependence on Galactic latitude weaker than in total
intensity. Synchrotron and dust emission can be expected to produce both
electric and magnetic polarization in similar
amounts~\cite{adc:zaldarriaga01eb}.

Synchrotron radiation is expected to be the dominant polarized foreground
at frequencies below $\sim 100\, \mathrm{GHz}$.
At the time of writing, the only well-surveyed
part of the sky for which polarized data is available is within the
Galactic plane, and then only at frequencies below 2.7 GHz
(e.g.~\cite{adc:duncan97}). The only
high-latitude observations available are very patchy~\cite{adc:brouw76}, and
most are not useful for assessing foreground contamination in clean regions
of the sky that will likely be the target
of future deep polarization observations. (But see~\cite{adc:bernardi03}
for a useful exception.) Analysis of the 2.7~GHz data of the Southern
Galactic plane from~\cite{adc:duncan97} gives $E$- and $B$-mode
power spectra going as $l^{-1.5}$~\cite{adc:giardino02},
but care should be taken in
extrapolating this to higher latitudes. A power law index of $-1.5$
is shallower than the large-angle polarization for both density
perturbations and gravitational waves.
Modelling at higher latitudes, based
on total intensity observations (e.g.~\cite{adc:giardino02}),
suggests that at frequencies
around $100\, \mathrm{GHz}$, synchrotron emission will not be a major
foreground for $E$-mode polarization, but that its removal will be essential
for large-angle $B$-mode searches.
We eagerly await the release of polarization maps
from the two-year WMAP data; these should greatly improve our knowledge of
high-latitude synchrotron polarization.

The power spectra from diffuse Galactic polarized dust emission have
been measured recently at $353 \, \mathrm{GHz}$ by the
Archeops team~\cite{adc:ponthieu05} over 20\% of the sky.
Extrapolating their measurements to
$100 \, \mathrm{GHz}$, they find for $3 \leq l \leq 70$ that
$(l+1) C_l^{E/B}/2\pi < 0.2 \, \mu \mathrm{K}^2$, excluding data with $5^\circ$
of the Galactic plane. Removal of the dust contamination is thus also likely
to be necessary for future large-angle $B$-mode searches.

A variety of methods have been developed to remove diffuse foreground
emission from CMB maps using multi-frequency data; see the contribution by
Delabrouille in this volume
for a review. Many of these were developed for total intensity data
(temperature), but they generalise to polarization straightforwardly
when applied directly to Stokes maps. 
Of these, some, such as Weiner filtering~\cite{adc:bouchet99},
require knowledge of the frequency spectra of all the components present, and
their angular power spectra.
Others attempt to derive this information from
the data by assuming statistical independence of the
components~\cite{adc:baccigalupi02,adc:delabrouille03}.
Finally, linear techniques, based on template fitting~\cite{adc:bennett03}, or
internal spectral combinations that preserve the CMB signal while minimising
the r.m.s.\ of the combined field, can also be applied very
effectively to polarization data provided that the spectral properties of the
contaminants do not vary too greatly over the survey area.
The results of applying these techniques to simulated polarization data
are very encouraging (e.g.~\cite{adc:tucci04}),
but we must await further observations to assess
the true extent to which foregrounds will compromise the scientific
returns of CMB polarization surveys.

Extra-galactic radio sources are a significant polarized contaminant
at small angular scales. They contribute equally to $E$- and $B$-mode
power and, for Poisson-distributed sources, have white angular power
spectra. An effective way to deal with these is simply to
exclude pixels within an observing beam of known sources above some flux
cut. Since the degree of polarization is typically rather low ($< 10 \%$),
source contamination for $E$-mode polarization is relatively less
troublesome than for temperature anisotropies. 
Having
Stokes maps with a large number of excluded pixels does complicate the
problem of separating $E$ and $B$ modes, and this is worse for low-resolution
observations since the fraction of the area that is removed is then larger.

\subsection{Map-level $E$ and $B$-mode separation}
\label{adc:secEB}

Separating the recovered polarization field into its electric and
magnetic parts amounts to solving for the potentials $P_E$ and $P_B$
in~(\ref{adc:eq3}). This can be done trivially if the observation covers
the full sky by using the orthogonality of the electric and magnetic tensor
harmonics to recover the multipoles $E_{lm}$ and $B_{lm}$ as
in~(\ref{adc:eq6}). However, problems arise for observations over only part
of the sky since the decomposition $\mathcal{P}_{ab} =
\nabla_{\langle a} \nabla_{b\rangle} P_E +
\epsilon^c{}_{\langle a}\nabla_{b\rangle} \nabla_c P_B$ is then not
unique~\cite{adc:lewis02,adc:bunn03}.
Even if fully-sky observations were available, we would want
to remove certain regions (e.g.\ the Galactic plane) from the subsequent
analysis, and this is most naturally done before the non-local decomposition
into $E$ and $B$ modes.

To see how the ambiguity comes about, consider attempting to solve for
$P_E$ and $P_B$ by recalling that the electric contribution to the
divergence $\nabla^b \clp_{ab}$ is curl-free and the magnetic part is
divergence free, i.e.
\begin{eqnarray}
2 \nabla^a \nabla^b \clp_{ab} &=& \nabla^2 (\nabla^2+2) P_E \; ,
\label{adc:eq26} \\
- 2 \epsilon_a {}_c \nabla^c \nabla^b \clp_{ab} &=&
\nabla^2 (\nabla^2+2) P_B \; .
\label{adc:eq27}
\end{eqnarray}
Solving these equations over part of the sphere requires one to specify
boundary data, namely the value of the potentials and their normal
derivatives on the boundary~\cite{adc:bunn03}.
But this information is not known since
it is the potentials that we are attempting to solve for.
We can, of course, impose
boundary conditions by force, in which case we should ensure that
we do not produce spurious $E$ or $B$ modes, i.e.\ we should recover
$P_B=0$ if there are no $B$-modes present. This will be true if we
solve~(\ref{adc:eq26}) and~(\ref{adc:eq27}) with the boundary conditions
that both the potentials and their gradients vanish on the
boundary~\cite{adc:lewis02,adc:bunn03}. In this manner,
we obtain a unique solution for $P_E$ and $P_B$, but they are systematically
wrong by $\Delta P_E$ and $\Delta P_B$ which are the unique solutions of
$\nabla^2 (\nabla^2 + 2) \Delta P_E = 0$ with $\Delta P_E = P_E$ and
$\nabla_a \Delta P_E = \nabla_a P_E$ on the boundary (similar equations
hold for $\Delta P_B$). Reconstructing the polarization field from the
recovered potentials, we see that this differs from the original field
by $\Delta \clp_{ab} \equiv \nabla_{\langle a} \nabla_{b\rangle} \Delta P_E +
\epsilon^c{}_{\langle a}\nabla_{b\rangle} \nabla_c \Delta P_B$. This
part of the polarization field, which satisfies
$\nabla^a \nabla^b \clp_{ab} = 0$ and $\epsilon_a {}_c \nabla^c
\nabla^b \Delta \clp_{ab}=0$, constitutes the ambiguous part since
it cannot be classified in terms of electric and magnetic modes. Some examples
of these ambiguous modes are illustrated in~\cite{adc:bunn03,adc:lewis03}.
A useful
vector analogy to keep in mind is that the ambiguous modes are like the
electromagnetic fields of (static) charges and currents that are localised
outside the observed region.

By excluding the ambiguous modes from subsequent analysis, we are throwing
away some information. As we might expect, the ambiguous modes tend to be
localised near the boundary, and a rough accounting of the fractional
loss of modes as a function of scale is given by the ratio of the
product of the boundary length and the coherence length to the observed
area.

\begin{figure}[t!]
\begin{center}
\includegraphics[height=9cm,angle=0]{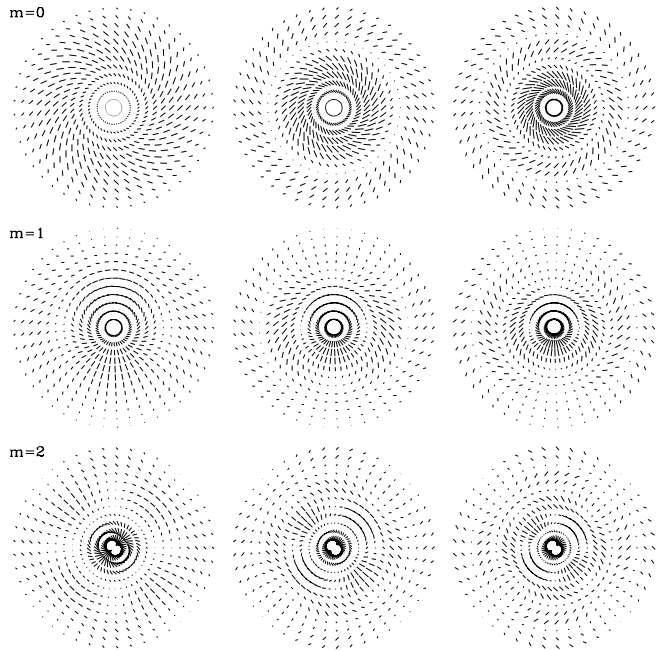}
\end{center}
\caption{Magnetic tensor basis functions $W_{B\, ab}$ for projecting onto pure
magnetic-polarization modes for an azimuthally-symmetric sky patch of radius
$10^\circ$~\cite{adc:lewis02}. Only $m=0$, 1 and 2 modes are shown.
In each case only the three modes with the largest expected power
are shown (with the left-most having greatest power).
}
\label{adc:fig8} 
\end{figure}

Practical methods for performing the separation described above
are given in~\cite{adc:lewis02,adc:bunn03,adc:lewis03}.
The basic idea is to avoid differentiating
the data by instead constructing tensor bases that can be used to
project out the unambiguous $E$ and $B$ modes. Consider, for example,
isolating magnetic polarization. If we construct a symmetric, trace-free
tensor $W_{B\, ab}  \equiv \sqrt{2}
\epsilon^c{}_{(a}\nabla_{b)}\nabla_c W$ from a scalar field $W$ that vanishes,
along with its gradient, on the boundary but is otherwise arbitrary.
Contracting with the polarization tensor and integrating by parts
gives~\cite{adc:lewis02}
\begin{equation}
\int  W_B^{ab\ast} \clp_{ab} \, \D \vnhat = - 2 \int
W^\ast \epsilon^b{}_c\nabla^c \nabla^a \clp_{ab} \, \D \vnhat \; ,
\label{adc:eq28}
\end{equation}
where there are no boundary terms on account of the conditions imposed
on $W$. It follows that the integral over the observed area is a non-local
measure of the $B$-mode component. Similarly, the tensor
$W_{E\, ab} \equiv \sqrt{2} \nabla_{\langle a} \nabla_{b \rangle} W$,
with the same scalar field $W$, extracts a measure of the $E$-mode
component. The construction of a complete set of such tensors can be performed
in spherical-harmonic space~\cite{adc:lewis02,adc:lewis03}
(see also Sect.~\ref{adc:sectpseudo}),
or directly in pixel space~\cite{adc:bunn03}.
An orthonormal tensor basis can be
constructed by taking the $W$ to be eigenfunctions of $\nabla^2 (\nabla^2 + 2)$
(with the boundary conditions given above). Eigenfunctions with different
eigenvalues are then orthogonal over the observed area, and the same is
true of the tensors derived from them. Examples of $W_{B\, ab}$ basis
functions for a circular patch of the sky are shown in Fig.~\ref{adc:fig8}.
In this case the tensor bases can be constructed with $W$ an eigenfunction
of $\partial / \partial_\phi$. For $m\ge 2$ exactly two $B$-modes are lost
per $m$; for $m=$ only one is lost; and for $m=0$ none are lost.
The basis functions in the figure are constructed to maximise the expected
power in that mode.

The loss of information due to the ambiguous modes has interesting
implications for survey design. For a fixed integration time, the `best'
strategy for measuring the variance of a Gaussian, statistically-isotropic
field, in the limit of low signal-to-noise, is to survey only a small field
since the error on the variance scales as the square-root of the survey area.
(Of course, this is not a wise strategy as you may be unlucky and choose
a field where the signal happens to be very low.) Applying this to
searches for gravitational waves with $B$-mode polarization, the implication
is that a small survey is best. However, the coherence length of the desired
signal is around a degree so the loss of information to ambiguous modes
becomes significant for small fields. The optimum field size on this basis
turns out to be a circular patch with radius around
$10^\circ$~\cite{adc:lewis02}. This argument ignores the effect of weak
gravitational lensing.
If we include the lens-induced $B$ modes
as an additional Gaussian noise term,\footnote{The $B$ modes produced by
lensing are not Gaussian. Their (connected) four-point function does
have a significant effect on the variance and covariance of the estimated
$B$-mode power spectrum~\cite{adc:smith04}.}
i.e.\ we make no attempt to clean out
the lensing contribution as described in Sect.~\ref{adc:seclensing},
the optimum size is increased further such that the contributions
to the error on the variance from thermal noise and lensing sample variance
are roughly equal.

We end this subsection by noting that coherent $E$-$B$ separation
is not strictly necessary for constraining cosmology. A statistical
separation can be attempted during power spectrum estimation
(see Sect.~\ref{adc:secpower}), and for the optimal maximum-likelihood
method this should be very effective. However, there are good reasons
to perform a coherent, map-level separation as well. For example,
visualising the $B$ modes will likely be a useful diagnostic of
unknown systematic effects and foreground residuals.

\subsection{Power spectrum estimation}
\label{adc:secpower}

The statistical properties of Gaussian random fields are fully specified
by their power spectra. For this reason, power spectrum estimation is
an important step in the analysis of CMB data. Ideally, this involves
mapping out the likelihood $\mathrm{Pr}(\vs | C_l^R)$,
where $\vec{s}$ are the observed CMB fields, as a function of the
power spectra $C_l^R$ ($R=T$, $E$, $B$ or $TE$).
In practice, what is normally done is to approximate
the likelihood as a Gaussian function of the power spectra, in which case
all of the information contained in the original CMB maps is distilled
into the position of maximum likelihood, and the curvature of the likelihood
there~\cite{adc:bond98}. In this process, we achieve a very significant
compression of the CMB maps into, at most, a few thousand maximum-likelihood
$C_l$s, and their associated covariances. Even this programme is
not achievable exactly for large data sets, so a number of faster techniques
based on estimators have been suggested. Here, the idea is to construct
(unbiased) estimators of the power spectra, and to use their covariances
-- or, better, their full sampling distributions~\cite{adc:wandelt01}
-- to constrain cosmology. A particularly fast class of estimators are
the heuristically-weighted, quadratic, or `pseudo-$C_l$'
estimators (e.g.~\cite{adc:hivon02,adc:szapudi01}).
For an interesting
critique of pseudo-$C_l$ and maximum-likelihood methods for temperature
anisotropies,
see~\cite{adc:efstathiou04}. For polarization, both methods can be applied
directly to maps of the Stokes parameters, in which case one must
rely on a statistical separation of the $E$ and $B$ modes, or after
map-level $E$-$B$ separation.

An important exception that demands a treatment beyond
the power spectrum is the effect of weak gravitational lensing. Special
techniques have been developed for this problem which allow reconstruction
of the underlying Gaussian CMB and deflection fields
(see Sect.~\ref{adc:seclensing}). The power spectra of these reconstructed
fields then contain all of the statistical (hence cosmological)
information present in the non-Gaussian, lensed fields.

\subsubsection{Maximum-likelihood methods}

Maximum-likelihood methods, and the problems one faces in their practical
implementation, are reviewed extensively elsewhere in this volume
by Borrill. Here, we briefly summarise the technique. Maximum-likelihood
methods have been used extensively in the analysis of real
polarization data, e.g. from DASI~\cite{adc:kovac02} and
CBI~\cite{adc:readhead04}.

For the direct analysis of Stokes maps (i.e.\ assuming no $E$-$B$ separation
has been performed) we combine the observed Stokes parameters in a
vector $\vs = (I_p , Q_p, U_p)^T$. The covariance matrix of this vector,
$\mC = \mS + \mN$, has contributions $\mS$ from the signal and $\mN$ from the
noise. The former is given essentially
by the temperature and polarization correlation functions (\ref{adc:eq11}),
but with the following modifications: (i) the power spectra are
multiplied by the appropriate product of the
beam functions $W_l$ and ${}_2 W_l$ to account for convolution with
the instrument beam; and (ii) there are additional factors to account for the
rotation of the Stokes parameters from the polar basis to the
geodesic basis defined by the points $p$ and $p'$. The noise contribution
$\mN$ includes the projection of the instrument noise onto the map,
which will generally be modified by the foreground removal process,
and an estimate of the covariance of any residual foreground errors.
Both $\mS$ and $\mN$ have $(3 N_\mathrm{pix})^2$ elements; $\mS$ is dense
in the pixel domain but is sparse in spherical-harmonic space, while
$\mN$ is typically sparse in the pixel domain. If map-level $E$-$B$ separation
has already been performed, the data $\vs$ can be taken as the unambiguous
$E$ and $B$ modes, and these can be analysed independently (although the
noise may correlate the errors in the estimated $E$- and $B$-mode power
spectra).
Assuming Gaussian signals
and errors, the likelihood function is
\begin{equation}
\mathrm{Pr}(\vs | C_l^R) = \frac{1}{\sqrt{\mathrm{det}(2\pi \mC)}}
\exp\left(- \vs^T \mC^{-1} \vs / 2\right) \, .
\label{adc:eq29}
\end{equation}
Maximising the likelihood is equivalent to maximising the posterior
probability $\mathrm{Pr}(C_l^R | \vs)$ if a uniform prior on the power spectra,
$C_l^R$, is adopted.

We now make some general remarks on this procedure.
\begin{itemize}
\item For full-sky coverage, and no noise, the
log-likelihood is (up to an irrelevant constant)
\begin{eqnarray}
-2 \ln \mathrm{Pr}(\vs | C_l^R) &=& \sum_l (2l+1)\left[\frac{1}{\Delta_l}
(\hat{C}_l^T C_l^E + \hat{C}_l^E C_l^T - 2 \hat{C}_l^{TE} C_l^{TE}) \right.
\nonumber \\
&&\mbox{} \phantom{xxxxxxxxxxxxxxx} \left.
+ \ln \Delta_l
+ \frac{\hat{C}_l^B}{C_l^B} + \ln C_l^B \right] \; ,
\label{adc:eqextra1}
\end{eqnarray}
where $\Delta_l \equiv C_l^T C_l^E - (C_l^{TE})^2$. The likelihood
is approximately a Gaussian function of the $C_l^R$ at high $l$.
The maximum-likelihood
point coincides with the simple estimators e.g.\ $\hat{C}_l^E = \sum_{m}
|E_{lm}|^2 / (2l+1)$ at low $l$.
\item Generally, the maximum of the likelihood can be located by solving
$\partial \ln \mathrm{Pr}(\vs | C_l^R) / \partial C_l^R = 0$,
with e.g.\ the Newton--Raphson method.
The data enters the
gradient of the log-likelihood, and hence the maximum likelihood point,
in the form $\vs^T \mC^{-1} (\partial \mS / \partial C_l^R) \mC^{-1} \vs$.
This is proportional to the simple estimate $\hat{C}_l^R$, but with the
spherical harmonic transforms taken only over the observed region, and with
the data first weighted non-locally by $\mC^{-1}$.
\item Approximating the likelihood as Gaussian requires the
computation of the curvature of the likelihood. 
At any $C_l^R$, the expectation value of the second derivatives of
(minus) the log-likelihood there,
under the assumption that these are the true power spectra,
defines the Fisher matrix. The inverse of the Fisher matrix,
evaluated at a smoothed version of the maximum likelihood power spectra, is
often quoted as the error on the power spectra. For the ideal case, where the
log-likelihood is given by~(\ref{adc:eqextra1}), the inverse of the Fisher
matrix has elements given by the cosmic variance formulae of
Sect.~\ref{adc:subseccosmic}.
\item A brute-force implementation to locate the maximum likelihood, and
evaluate the Fisher matrix, is  not feasible for mega-pixel data sets.
Both the operation count, $O(N_{\mathrm{pix}}^3)$, and the storage requirements
are prohibitive.
\item Additional regularisation is required for small surveys since
the problem is then degenerate if we attempt to solve for every multipole $l$.
A common fix is to solve for bandpower amplitudes~\cite{adc:bond98}. For narrow
bands, approximating the $l(l+1)C_l$s as piecewise constant is common. As the
polarization power spectra are steeper functions of $l$ than $C_l^T$,
particularly on large scales, shaped bandpowers are often
employed in polarization analyses (e.g.~\cite{adc:kovac02}).
\end{itemize}

Maximum-likelihood estimation is closely related to the optimal
quadratic estimators~\cite{adc:tegmark97,adc:tegmark01}.
These take the form
\begin{equation}
\hat{C}_l^R = \frac{1}{2} \sum_{l' R'} F^{-1}_{(lR)(l'R')} \mathrm{trace}
[(\vs \vs^T - \mN)\mC^{-1} \partial \mS/\partial C_{l'}^{R'}\mC^{-1}],
\label{adc:eq30}
\end{equation}
where the Fisher matrix is $F_{(lR)(l'R')} \equiv
\mathrm{trace}[\mC^{-1} \partial \mS/\partial
C_l^R \mC^{-1} \partial \mS/\partial C_{l'}^{R'}]/2$. The quadratic estimator
is unbiased for any (invertible) matrix $\mC$, but the variance is minimised
if $\mC$ is truly the covariance of the data. Of course, this is
not known a priori so we instead construct an approximation to $\mC$
from some reasonable guess for the power spectra. It can be shown that
iterating the quadratic estimator (i.e.\ constructing $\mC$ for the next
estimate from the current power spectrum estimates $\hat{C}_l^R$)
is equivalent to performing a Newton--Raphson maximisation of the
likelihood, with the curvature approximated by the Fisher
matrix~\cite{adc:bond98}.
The optimal quadratic estimator has the same computational problems
as maximum likelihood methods.

\subsubsection{Heuristically-weighted, quadratic estimates}
\label{adc:sectpseudo}

The computational complexity of the optimal quadratic
estimator~(\ref{adc:eq30}) arises from the non-local weighting of the
data by the inverse covariance matrix $\mC^{-1}$. Heuristically-weighted,
quadratic
methods~\cite{adc:chon04,adc:wandelt01,adc:hivon02,adc:szapudi01,adc:hansen03}
circumvent this problem by adopting a local
weighting scheme. The use of
fast-spherical transforms, as implemented in e.g.\ the HEALPIX
package~\cite{adc:gorski04},
reduce the operations count to $O(N_{\mathrm{pix}}^{3/2})$
which is fast enough to be used in high-volume simulations.

We shall concentrate here on the polarization auto-correlations only.
Including the cross-correlation between temperature and polarization is
straightforward~\cite{adc:chon04,adc:hansen03}.
Assuming we are analysing Stokes maps, we adopt a
real weighting $w(\vnhat)$ that is zero where we have no data. For the
case of temperature anisotropies, weighting with the inverse noise variance
on scales where the noise dominates the signal, and uniformly where the
signal dominates, can be shown to be close to optimal~\cite{adc:efstathiou04}.
Care should be taken in generalising this reasoning to
low-amplitude $B$-mode polarization because of the additional
complication of $E$-$B$ mixing. We extract the electric
and magnetic multipoles of the weighted polarization field with
spherical transforms,
\begin{eqnarray}
\tilde{E}_{lm} \pm \I \tilde{B}_{lm} &\equiv& \int w(\vnhat) (Q\mp \I U)
(\vnhat) {}_{\pm 2}Y_{lm}^*(\vnhat) \, \D \vnhat \; ,\nonumber \\
&=& \sum_{(lm)'} {}_{\pm 2} I_{(lm)(lm)'}(E_{(lm)'} \pm \I B_{(lm)'}) \; ,
\label{adc:eq31} 
\end{eqnarray}
where the Hermitian coupling matrices are
\begin{equation}
{}_{\pm 2} I_{(lm)(lm)'} \equiv \int w(\vnhat) {}_{\pm 2} Y_{(lm)'}(\vnhat)
{}_{\pm 2} Y_{lm}^*(\vnhat) \, \D \vnhat \; ,
\label{adc:eq32}
\end{equation}
and can be expressed in terms of the (scalar) multipoles, $w_{lm}$, of the
weight function, and the Wigner $3j$ symbols~\cite{adc:hansen03}.
We have left the effect of smoothing by the instrument beam implicit
in~(\ref{adc:eq31}); the factors ${}_2 W_l$ can be absorbed into the
multipoles $E_{lm}$ and $B_{lm}$, and, later, ${}_2 W_l^2$ into the power
spectra $C_l^E$ and $C_l^B$.
We compress the pseudo-multipoles into \emph{pseudo-$C_l$s}
according to
\begin{equation}
\tilde{C}_l^E \equiv \frac{1}{2l+1} \sum_m |\tilde{E}_{lm}|^2 \quad , \quad
\tilde{C}_l^B \equiv \frac{1}{2l+1} \sum_m |\tilde{B}_{lm}|^2 \; .
\label{adc:eq33}
\end{equation}
These quantities are equivalent to the
$\vs^T \mC^{-1} (\partial \mS / \partial C_l^R) \mC^{-1} \vs$ that
appear in the optimal quadratic estimator~(\ref{adc:eq30}) if we replace
$\mC^{-1}$ by heuristic weights.

\subsubsection{$E$-$B$ mixing revisited}

We make a brief aside to note the relation between the pseudo-multipoles
and the issue of separating $E$ and $B$ modes with incomplete sky coverage.
The pseudo-multipoles mix $E$ and $B$ modes according to
\begin{eqnarray}
\tilde{E}_{lm} &=& \sum_{(lm)'} 
\left( {}_+ I_{(lm)(lm)'}E_{(lm)'}+\I {}_{-}I_{(lm)(lm)'}B_{(lm)'} \right) \; ,
\label{adc:eq34}\\
\tilde{B}_{lm} &=& \sum_{(lm)'}
\left( {}_+ I_{(lm)(lm)'}B_{(lm)'} - \I {}_{-}I_{(lm)(lm)'}E_{(lm)'}\right)\; ,
\label{adc:eq35}
\end{eqnarray}
where we have defined the Hermitian ${}_\pm I_{(lm)(lm)'}$ to be
\begin{eqnarray}
{}_+ I_{(lm)(lm)'} &\equiv& \frac{1}{2}({}_{+2} I_{(lm)(lm)'}
+ {}_{-2}I_{(lm)(lm)'}) = \int w Y^{E\, *}_{(lm)\, ab} Y^{E\, ab}_{(lm)'}
\, \D \vnhat \; , \label{adc:eq36} \\
{}_{-}I_{(lm)(lm)'} &\equiv& \frac{1}{2}({}_{+2}I_{(lm)(lm)'} -
{}_{-2}I_{lm(lm)'}) = -i \int w Y^{E\, *}_{(lm)\, ab} Y^{B\, ab}_{(lm)'}
\, \D \vnhat\; .
\label{adc:eq37}
\end{eqnarray}
The mixing matrix ${}_- I_{(lm)(lm)'}$
controls the mixing of $E$ modes into $\tilde{B}_{lm}$, and vice versa.
For the case of uniform weighting over the full sky, the mixing
matrix vanishes and ${}_{+} I_{(lm)(lm)'} \propto \delta_{ll'} \delta_{mm'}$.
For partial sky coverage,
if $w(\vnhat)$ is unity inside the observed region but zero elsewhere,
integrating by parts
in~(\ref{adc:eq37}) shows that ${}_{-}I_{(lm)(lm)'}$ reduces to a line
integral around the boundary~\cite{adc:lewis02}:
\begin{equation}
i {}_{-}I_{(lm)(lm)'} = \sqrt{\frac{2(l-2)!}{(l+2)!}}
\oint \left(\nabla_b Y^*_{lm} Y^{B\, ab}_{(lm)'} - Y^*_{lm}
\nabla_b Y^{B\, ab}_{(lm)'} \right) \, \D l_a \; .
\label{adc:eq38}
\end{equation}
If we contract the $\tilde{B}_{lm}$ with the multipoles $W^*_{lm}$
of any scalar
function $W(\vnhat) = \sum_{lm} W_{lm} Y_{lm}(\vnhat)$, such that
$W$ and $\nabla_a W$ vanish on the boundary of the observed region, we see
that $\sum_{lm} W^*_{lm} {}_{-}I_{(lm)(lm)'}=0$ and we will
have projected out the $E$-mode contamination from the $\tilde{B}_{lm}$.
Since the construction
$\sum_{lm} W^*_{lm} \tilde{B}_{lm}$ is equivalent to integrating the
contraction of $W_B^{ab\, *} = \sqrt{2} \epsilon^c{}_{(a}\nabla_{b)}\nabla_c
W^*$ and $\clp_{ab}$ over the observed region, isolating those
functions in harmonic space that are orthogonal to the range of
$ {}_{-}I_{(lm)(lm)'}$ is equivalent to finding elements of the
tensor basis of the unambiguous $E$ and $B$ modes over the observed region
(see Sect.~\ref{adc:secEB}). The harmonic-space method for
$E$-$B$ separation developed in~\cite{adc:lewis02}
uses singular value techniques to
project out the range of the mixing matrix ${}_{-}I_{(lm)(lm)'}$.

For a more general weight function $w(\vnhat)$, if we assume that it
smoothly apodizes the edges of a connected region of linear dimension
$\sim R$, the relative sizes of ${}_{-}I$ and ${}_{+}I$ at scale $l^{-1}$
are in the ratio $1/(lR)$~\cite{adc:challinor04}.
It follows that the geometric effect of mode mixing
is suppressed on scales small compared to the survey size.

\subsubsection{Mean values of the pseudo-$C_l$s}

\begin{figure}
\begin{center}
\includegraphics[width=7cm,angle=-90]{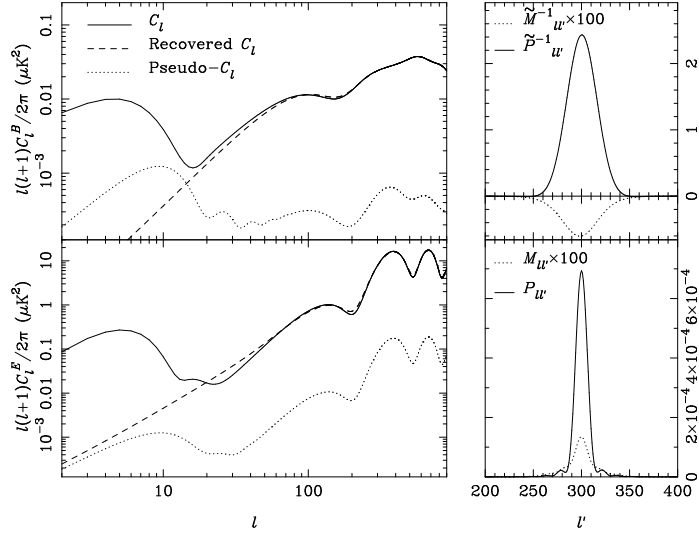}
\end{center}
\caption{Left: power spectra for $B$ (top) and $E$ (bottom; solid lines)
compared to the mean pseudo-$C_l$s (dotted lines) and the recovered power
spectra (dashed lines). Right: the bottom panel shows representative
matrices $P_{ll'}$ (solid lines) and $M_{ll'}$ (dotted lines);
the top panel shows the pseudo-inverses
$\tilde{P}^{-1}_{ll'}$ (solid lines) and $\tilde{M}^{-1}_{ll'}$
(dotted lines). Note that
$M_{ll'}$ and $\tilde{M}^{-1}_{ll'}$ have been multiplied by a factor of
100 for clarity. The weight function applied to the map is uniform inside
a circle of $10^\circ$ radius, with cosine apodization out to $15^\circ$.
To obtain the pseudo-inverses, a Gaussian apodization of $4^\circ$ HWHM is
applied to the correlation functions.
\label{adc:fig9}
}
\end{figure}

Returning to the pseudo-$C_l$s, their mean values contain two
contributions, one from the covariance of the instrument noise
(and foreground residuals) and one from the CMB power spectra. The former
can be estimated by using a large number of simulated observations:
$\sim 10^4$ observations should give $\sim 1\%$ accuracy. The mean level
in the simulations can then be subtracted from the observed $\tilde{C}_l$
to remove the noise bias in the mean. From now on we shall assume that
the pseudo-$C_l$s have been corrected in this way. Alternatively,
if only correlations between different instrument channels are used in the
construction of the pseudo-$C_ls$, the noise bias will vanish if the noise
is uncorrelated between channels~\cite{adc:hinshaw03}.
The means of the
pseudo-$C_l$s are then linearly related to the true power spectra:
\begin{equation}
\langle \tilde{C}_l^E \rangle = \sum_{l'} P_{ll'} C_{l'}^E + M_{ll'} C_{l'}^B
\; , \qquad \langle \tilde{C}_l^B \rangle = \sum_{l'}
M_{ll'} C_{l'}^E + P_{ll'} C_{l'}^B \; .
\label{adc:eq39}
\end{equation}
Here, the matrices are $P_{ll'} \equiv \sum_{m m'} | {}_+ I_{(lm)(lm)'}|^2
/(2l+1)$, with a similar expression for $M_{ll'}$. They depend on
the weight function $w(\vnhat)$ only through the rotational invariant
$w_l \equiv \sum_m | w_{lm} |^2 / (2l+1)$, and can be expressed in terms
of $3j$ symbols as
\begin{equation}
(P/M)_{ll'} = \frac{2l'+1}{8\pi} \sum_L (2L+1) w_L [1 \pm
(-1)^{l+l'+L}] \left( \begin{array}{ccc} l & l' & L \\ -2 & 2 & 0 \end{array}
\right)^2 \; .
\label{adc:eq40}
\end{equation}
For uniform weighting over the full sky, only $w_0$ is non-zero and
so $M_{ll'}$ vanishes as it must (the $3j$ symbol forces $l=l'$ and
hence $l+l'+L$ to be even). For weight functions that smoothly apodize
observations over a region of linear size $\sim R$, the weight function
is almost band-limited to $l_{\mathrm{max}} \sim 1/R$. In this case the
$P_{ll'}$ and $M_{ll'}$ matrices are close to band diagonal, and, for
spectra that are smooth on the scale of $R^{-1}$, the means of the
pseudo-$C_l$s are approximately
\begin{equation}
\langle \tilde{C}_l^E \rangle \approx C_l^E \sum_{l'} P_{ll'} + C_l^B
\sum_{l'} M_{ll'} \; , \qquad
\langle \tilde{C}_l^B \rangle \approx C_l^E \sum_{l'} M_{ll'} + C_l^B
\sum_{l'} P_{ll'} \; . 
\label{adc:eq41}
\end{equation}
The relative size of the mixing of $E$- and $B$-mode power in the
pseudo-$C_l$s is thus set by the ratio of the normalisations
$\sum_{l'} M_{ll'}$ to $\sum_{l'} P_{ll'}$. For $l \gg 1/R$, it can
be shown that~\cite{adc:challinor04}
\begin{eqnarray}
\sum_{l'} P_{ll'} &\approx& \frac{1}{4\pi} \int w^2(\vnhat) \, \D\vnhat \; ,
\label{adc:eq42} \\
\sum_{l'} M_{ll'} &\approx& \frac{1}{2\pi} \frac{1}{l(l+1)}
\int (\nabla w)^2(\vnhat) \, \D\vnhat \; ,
\label{adc:eq43}
\end{eqnarray}
so that mixing of power is suppressed by $1/(lR)^2$. Given that it is
expected that $C_l^E \gg C_l^B$ on all scales, $\langle \tilde{C}_l^E
\rangle$ is essentially a scaled version of the true spectrum, but
this need not be the case for $\tilde{C}_l^B$. These ideas are illustrated
in Fig.~\ref{adc:fig9} which show typical rows of the matrices $P_{ll'}$
and $M_{ll'}$, and the mean pseudo-$C_l$s that they give rise to, for
observations  covering a circle of radius $15^\circ$ with the last
$5^\circ$ apodized with a cosine function. The mean of $\tilde{C}_l^E$
is seen to be a scaled version of the true power spectrum except on large
scales, but this is not the case for $\tilde{C}_l^B$.

\subsubsection{`Unbiased' estimators}

The pseudo-$C_l$s can be used directly for parameter
estimation~\cite{adc:wandelt01},
but it is often desirable to deconvolve the
geometric effects of the weight function so we have unbiased estimates
of the underlying power spectra.
For example, for presentation
purposes it is useful if we can plot the data in a form where they can be
easily compared with true power spectra.

For observations covering a
significant fraction of the sky, deconvolution is straightforward. The
matrices $P_{ll'} \pm M_{ll'}$ will then be invertible and we can form
unbiased estimates of the power spectra according to
\begin{equation}
\hat{C}_l^E \pm \hat{C}_l^B = \sum_{l'} (P\pm M)^{-1}_{ll'}(\tilde{C}_{l'}^E
\pm \tilde{C}_{l'}^B) \; . \label{adc:eq44}
\end{equation}
An equivalent formulation of this approach can be made in terms of correlation
functions~\cite{adc:chon04,adc:szapudi01}. Consider estimating the correlation
functions~(\ref{adc:eq10}) by averaging over pairs of pixels separated
by $\beta$, with pairs weighted by $w(\vnhat_1)w(\vnhat_2)$. It turns out
that the data enters the estimators only through its pseudo-$C_l$s. The
estimators $\hat{\xi}_\pm (\beta)$ for $\langle (\bar{Q} \mp \I \bar{U})
(\vnhat_1) (\bar{Q} + \I \bar{U})(\vnhat_2) \rangle$, where
$\vnhat_1 \cdot \vnhat_2 = \cos\beta$, can be expressed as~\cite{adc:chon04}
\begin{equation}
\hat{\xi}_\pm (\beta) = \frac{\sum_l (2l+1)
(\tilde{C}_l^E \pm \tilde{C}_l^B)
d^l_{2\, \pm 2}(\beta)}{\sum_{l} w_l P_l(\cos\beta)} \; ,
\label{adc:eq45}
\end{equation}
which also provides an efficient $O(N_{\mathrm{pix}}^{3/2})$ way of
estimating the correlation functions. If enough sky is available to
estimate the correlation functions for all $\beta$, then we can recover
unbiased estimates of the power spectra by integral transforms. Using
the orthogonality of the reduced Wigner functions, we have
\begin{equation}
\hat{C}^E_l \pm \hat{C}^B_l = 2\pi \int_{-1}^{+1} \hat{\xi}_\pm (\beta)
d^l_{2\, \pm 2} (\beta) \, \D\cos\beta \; .
\label{adc:eq46}
\end{equation}
The integration can be performed essentially exactly with
Gauss--Legendre integration, and a significant feature of estimating
the correlation functions via the pseudo-$C_l$s is that they can be evaluated
for any $\beta$ up to the maximum pixel separation in the map. This avoids
the need to re-sample the correlation functions onto the points required
for the quadrature. This method has recently been applied to
estimate the power spectra from the polarized channel of
Archeops~\cite{adc:ponthieu05}.
Aside from discretisation issues, this correlation function method is
equivalent to performing the direct inversion in~(\ref{adc:eq44}).

For small surveys we will not have pixels separated by all angles
$\le 180^\circ$, in which case we cannot estimate the correlation functions
for all $\beta$. The inversion in~(\ref{adc:eq46}) is then not possible.
This is equivalent to saying that the survey does not have the spectral
resolution to perform the matrix inversions in~(\ref{adc:eq44}).
Various schemes can be adopted to regularise the inversion, such as
the use of bandpowers~\cite{adc:hivon02}. Alternatively, we can
settle for pseudo-inverses $\tilde{P}^{-1}_{ll'}$ and
$\tilde{M}^{-1}_{ll'}$, constructing estimators, e.g.\
\begin{equation}
\hat{C}_{l}^{B} = \sum_{l'} \tilde{P}^{-1}_{ll'} \tilde{C}_{l'}^{B}
+ \tilde{M}^{-1}_{ll'} \tilde{C}_{l'}^{E} \; ,
\label{adc:47}
\end{equation}
that have a simpler relation to the true power spectra than do the
pseudo-$C_l$s. One desirable property we might enforce on the
pseudo-inverses is that $E$-$B$ mixing is removed \emph{in the mean},
i.e.\ $\langle \hat{C}_l^B \rangle$ has no contribution from
$C_l^E$. A convenient way to find such a pseudo-inverse is given
in~\cite{adc:chon04}, which builds on earlier work in the context of
cosmic shear analysis~\cite{adc:crittenden02}. We make use
of the correlation functions estimates~(\ref{adc:eq45}) in the angular
range $(0,\beta_{\mathrm{max}})$, where $\beta_{\mathrm{max}}$ is the maximum
separation of pixels in the Stokes maps. It is shown in~\cite{adc:chon04}
that the estimators $\hat{\xi}(\beta) \pm \hat{\xi}_-(\beta)$, which
are linear in the pseudo-$C_l$s, contain only $E$ and $B$-mode power,
respectively, in the mean. The function $\hat{\xi}(\beta)$ is obtained
in the range $(0,\beta_{\mathrm{max}})$ by quadrature of $\hat{\xi}_+$
in the same range:
\begin{eqnarray}
\hat{\xi}(\beta) &=& \hat{\xi}_+(\beta) +
\frac{1}{\sin^2(\beta/2)}\int_{\cos\beta}^1
 \xi_+(\beta') \sec^4(\beta'/2) \, \D \cos\beta' \nonumber \\
&&\mbox{} -
\frac{2(2+\cos\beta)}{\sin^4(\beta/2)} \int_{\cos\beta}^1
\xi_+(\beta') \frac{\tan^3(\beta'/2)}{\sin\beta'} \, \D \cos\beta' \; ,
\label{adc:eq48}
\end{eqnarray}
and satisfies
\begin{equation}
\frac{1}{2} \langle \hat{\xi}(\beta) \pm \hat{\xi}_- (\beta) \rangle
= \sum_l \frac{2l+1}{4\pi} C_l^{E/B} d^l_{2\, -2}(\beta) \; .
\label{adc:eq49}
\end{equation}
We can recover estimates $\hat{C}_l^E$ and $\hat{C}_l^B$ that are
linear combinations of the true power spectra, but with no mixing of
$E$ and $B$ in the mean, by an apodized integral transform:
\begin{equation}
\hat{C}^{E/B}_l = 2\pi \int_{\cos\beta_{\mathrm{max}}}^1
\frac{1}{2} [\hat{\xi}(\beta) \pm \hat{\xi}_-(\beta)]
d^l_{2\, -2}(\beta) f(\beta) \, \D \cos\beta \; ,
\label{adc:eq50}
\end{equation}
where $f(\beta)$ is chosen to apodize the discontinuity at
$\beta_{\mathrm{max}}$. In this manner, we construct pseudo-inverses
that deconvolve the pseudo-$C_l$s up to a window function
${}_{-2} K_{ll'}$, i.e.\
$\langle \hat{C}_l^E \rangle = \sum_{l'} {}_{-2} K_{ll'} C_{l'}^E$ and
similarly for $C_l^B$. The window function is determined solely by
the apodizing function $f(\beta)$ and is given by
\begin{equation}
{}_{-2} K_{ll'} = \frac{2l'+1}{2} \int_{\cos\beta_{\mathrm{max}}}^1
f(\beta) d^l_{2\, -2}(\beta) d^{l'}_{2\, -2}(\beta) \, \D \cos\beta \; .
\label{adc:eq51}
\end{equation}
Oscillations in the window function can be suppressed with a careful
choice of apodizing function, but the minimum achievable width
is determined by the survey size $1/ \beta_{\mathrm{max}}$.
An example of the construction of `unbiased estimates' with this
route is given in Fig.~\ref{adc:fig9}.

\subsubsection{Covariance of the pseudo-$C_l$s}

The covariance matrix of the pseudo-$C_l$s can be used to quantify the
errors on any heuristically-weighted quadratic estimate of the power
spectra. Having accurate error information is clearly essential to establish
reliable constraints on cosmological models.
For Gaussian fields, computing the pseudo-$C_l$ covariance directly is
simple in principle, but the computational costs are prohibitive at
high $l$. This is particularly problematic given that the sample
covariance (i.e.\ that due to the random nature of the CMB fields)
depends on the power spectra, and so should ideally be recomputed at
each point in parameter space when constraining models.
This model dependence also makes computing the sample variance `blindly'
from a large-suite of simulations unattractive.
Fortunately, for the temperature anisotropies the sample
covariance can be accurately approximated with analytic methods for $l$ large
compared to the inverse of the survey size~\cite{adc:efstathiou04}.
Extending this calculation to polarization is difficult because the effect
of $E$-$B$ mixing can be the dominant source of variance in
$\tilde{C}_l^B$ on a wide range of scales for small surveys. However,
useful analytic approximations do now exist~\cite{adc:challinor04}.
For an accurate calculation on all scales, one can combine the analytical
calculations at high $l$ (possibly calibrated from a small number of
simulations if required) with a direct calculation at reduced resolution at low
$l$~\cite{adc:efstathiou04}. (See also~\cite{adc:brown04} for an alternative
prescription.) The noise contribution (and any residual errors from
foreground removal) is generally best handled with simulations.

\begin{figure}[t!]
\begin{center}
\includegraphics[width=7cm,angle=-90]{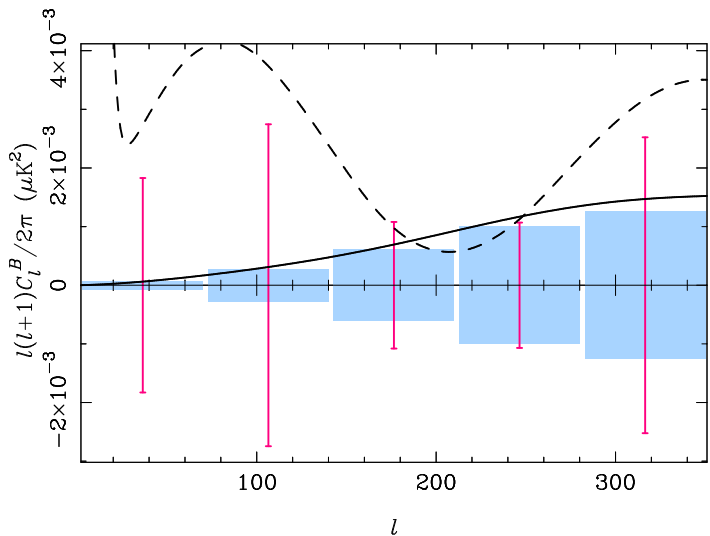}
\end{center}
\caption{Sample variance errors on the recovered $\hat{C}_l^B$ using the
estimator~(\ref{adc:eq50}) in the null hypothesis $r=0$.
The survey area and weight function are the same as in 
Fig.~\ref{adc:fig9}.
The error boxes are the contribution purely from $C_l^B$ (i.e.\
the lens-induced $B$ modes) to the
one-sigma errors on flat bandpower estimates with a $\Delta l = 70$. These
are thus representative of the errors that would be obtained with
the pseudo-$C_l$ method if it were applied to maps from which the $E$ modes
had been removed (see Sect.~\ref{adc:secEB}, but here we have ignored the loss
of information to the ambiguous modes).
The error bars are the contribution purely from $C_l^E$ and arise from
$E$-$B$ mixing. The dashed and solid lines are crude `rule-of-thumb'
approximations to the errors; see~\cite{adc:challinor04} for full details.
\label{adc:fig10}
}
\end{figure}

An important issue for heuristically-weighted quadratic estimators is that
the weighting scheme adopted generally makes no attempt to separate $E$ and
$B$ modes coherently. While we can remove $E$-$B$ mixing from the
power spectra in the mean, in any realisation $\hat{C}_l^B$ need not vanish
even if there are no $B$-modes present. This cross-contribution to the sample
variance of pseudo-$C_l$ estimators makes them inaccurate for high-sensitivity,
small-area $B$-mode surveys. This is illustrated in Fig.~\ref{adc:fig10},
where it is shown that the contribution to the sample variance
of $\hat{C}_l^B$ due to leakage of $E$ modes is well in excess of that due to
lens-induced $B$ modes. (We have assumed there are no gravitational waves
present, so lensing is the only source of $B$ modes.)
For the $15^\circ$-radius survey considered,
with no instrument noise, this excess variance increases the error on the
tensor-to-scalar ratio $r$ in the null hypothesis to $\Delta r=0.15$
($3 \sigma$).
One way to remove this problem is first to isolate the $B$ modes with the
methods described in Sect.~\ref{adc:secEB}. Ignoring the loss in this process,
we find that the limit on $r$ is then
reduced to $3.0 \times 10^{-4}$; this arises
solely from the sample variance of the lens-induced $B$ modes.

\subsection{Non-Gaussianity and lensing reconstruction}
\label{adc:seclensing}

The effect of lensing
on the CMB will probably be seen first in the angular power
spectra (e.g.~\cite{adc:zaldarriaga98lens,adc:seljak96}
and references therein),
or, indirectly, via cross-correlation with large-scale
structure~\cite{adc:hirata04}.
As with gravitational-wave searches, the most promising place to look
from the standpoint of cosmic variance is the $B$-mode spectrum
since, on small scales, lensing may be the dominant signal. Lensing
is sensitive to the evolution of the gravitational potential at low
redshift and can thus can be used to probe parameters, such as neutrino
masses and the properties of dark energy, to which the primary CMB spectra are
insensitive~\cite{adc:hu02synergy}.
At the sensitivity of the Planck experiment,
the effect of lensing on the observed power spectra must be included to
avoid bias when estimating parameters such as the baryon
density. As thermal noise levels improve, and the
sample variance of lens-induced $B$-modes comes to dominate the error budget
on $C_l^B$, corrections due to the non-Gaussianity from lensing will
need to be included in the $C_l^B$ errors and their
covariance~\cite{adc:smith04}.

Looking further ahead, a number of authors have proposed reconstructing the
lensing deflection field directly from the CMB by exploiting the
non-Gaussianity of the lensing effect~(see
e.g.~\cite{adc:seljak99lensrecon,adc:hu01lensrecon} and
references therein). This allows one
to extract more information from the observed (lensed) CMB fields than is
possible with an analysis of their power spectra alone, as well as providing
a coherent reconstruction of the projected mass distribution.
In essence, the reconstruction methods work by using the locally-anisotropic
effect of lensing shear on the statistics of the CMB fluctuations
within a region over which the deflections are coherent ($\sim 60^\circ$).
The fact that we do not know the unlensed CMB fields, but only their
statistical properties, introduces a noise in the reconstruction analogous
to the effect of the scatter in intrinsic ellipticities of background
galaxies in cosmic shear analyses. CMB polarization is particularly useful
here~\cite{adc:benabed01,adc:hu02lensrecon,adc:hirata03}:
it has more power on small
scales than the temperature
anisotropies and so allows reconstruction of the deflection to smaller
scales. It was shown in~\cite{adc:hu02lensrecon} that \emph{local}
correlations between $E$ and the lens-induced $B$ modes, estimated with
simple quadratic estimators, reconstruct the deflection field with
the highest signal-to-noise on all scales once thermal noise permits
imaging of the lens-induced $B$ modes on small scales. This requires
both high sensitivity ($\sim 1 \, \mu\mathrm{K}$ in an arcmin pixel),
and also high angular resolution (a few arcmin). The $E$-$B$ estimator
is particularly powerful since there is no contribution to its sample
variance from chance correlations in the unlensed fields since
there are no unlensed small-scale $B$ modes.

Recently, it has been argued that the relatively larger non-Gaussianity
of the small-scale (lensed) $B$-modes, compared to the temperature
anisotropies and $E$ modes, allows further improvement in the lensing
reconstruction if more optimal techniques are employed~\cite{adc:hirata03}.
Indeed, a simple counting argument suggests that, in the absence of thermal
noise, the assumption that lensing is the only source of
small-scale $B$ modes allows perfect reconstruction of the deflection
field (except possibly for a few degenerate configurations). In practice,
errors in the theoretical modelling of lensing ultimately limit the
reconstruction, but this effect is unimportant until
thermal noise levels fall well below the very optimistic level of
$0.25\, \mu$K-arcmin.

An important application of lensing reconstruction is to `clean out'
the large-scale $B$ modes where any gravitational wave background is
expected to contribute. If the lens-induced modes are treated simply
as an additional
source of Gaussian confusion, the limit on the detectability imposed by
lensing (i.e.\ with no thermal noise or errors in foreground removal)
is $r > 4 \times 10^{-5}$ at $3\sigma$ for a full-sky survey~\cite{adc:knox02}.
This can be improved by more than a factor of 40 with optimal reconstruction
methods~\cite{adc:seljak04lens},
corresponding to an energy scale of inflation $V^{1/4}
\approx 1 \times 10^{15}\, \mathrm{GeV}$.

\section{Conclusion}
\label{adc:secconclusion}

Detections of CMB polarization are still at an early stage, but already
we are beginning to see the promise of polarization data for constraining
cosmological models being realised. The electric-polarization power
spectrum measurements from DASI~\cite{adc:leitch04} and
CBI~\cite{adc:readhead04} reveal
acoustic oscillations with an amplitude and phase perfectly
consistent with the best-fit adiabatic models to the temperature anisotropies.
This is a non-trivial test on the dynamics of the photons and baryons
around the time of recombination. Furthermore, the large-angle measurements
of the temperature--polarization cross-correlation from WMAP~\cite{adc:kogut03}
indicate a large optical depth to reionization and hence a complex ionization
history. We can expect further rapid progress observationally, with
more accurate measurements of the power spectra of
$E$-mode polarization, and its correlation with the temperature anisotropies,
expected shortly from a number of ground and balloon-borne experiments.
These experiments will also greatly
increase our knowledge of polarized astrophysical foregrounds at CMB
frequencies. Instruments are currently being commissioned
that should have the sensitivity to detect the power of $B$-mode
polarization induced by weak gravitational lensing~\cite{adc:bowden04},
and, already,
several groups are working on a new generation of polarimeters with
the ambition of detecting gravitational-waves and reconstructing the
projected mass distribution from CMB polarization observations.

The success of these programmes will depend critically on many complex
data analysis steps. We have attempted to summarise here some of the generic
parts of a polarization data analysis pipeline, but, inevitably, have had to
leave out many topics that are more instrument-specific. Important
omissions include calibration of the instrument, cleaning and other
low-level reductions of time-stream data, noise estimation,
propagation of errors, and the broad topic of statistical accounting
for non-ideal instrument effects. Given the exquisite
control of systematic effects -- both instrumental and
astrophysical -- that searching for sub-$\mu\mathrm{K}$ signals
demands~\cite{adc:hu03systematic},
these omissions will almost certainly prove to be the most critical steps.

\section*{Acknowledgments}

AC acknowledges a Royal Society University Research Fellowship and thanks
the organisers for the invitation to participate in
an interesting and productive summer school.

%
%
%



\end{document}